\documentclass[aps, prd, letterpaper, twocolumn, amsmath, amssymb, preprintnumbers, superscriptaddress, floatfix, nofootinbib, longbibliography]{revtex4-1}

\usepackage[T1]{fontenc}
\usepackage{graphicx}
\usepackage{amsmath}
\usepackage{amsfonts}
\usepackage[colorlinks=true, linkcolor=blue, urlcolor=blue, citecolor=blue, anchorcolor=blue]{hyperref}
\usepackage{bbold}

\newcommand{\ignore}[1]{}
\newcommand\SU{\mathrm{SU}}
\newcommand\U{\mathrm{U}}
\newcommand\SO{\mathrm{SO}}
\newcommand{\cL}{\ensuremath{\mathcal L} }
\newcommand{\cM}{\ensuremath{\mathcal M} }
\newcommand{\cO}{\ensuremath{\mathcal O} }
\newcommand{\Tr}[1]{\ensuremath{\mbox{Tr}\left[ #1 \right]} }
\newcommand{\vev}[1]{\ensuremath{\left\langle #1 \right\rangle} }
\newcommand{\eq}[1]{Eq.~(\ref{#1})}
\newcommand{\fig}[1]{Fig.~\ref{#1}}
\newcommand{\refcite}[1]{Ref.~\cite{#1}}
\newcommand{\secref}[1]{Section~\ref{#1}}

\newcommand{\bottomrule}{\hline}
\newcommand{\midrule}{\hline}
\newcommand{\la}{\left\langle}
\newcommand{\ra}{\right\rangle}

\begin{document}
\preprint{RIKEN-iTHEMS-Report-21}
\preprint{LLNL-JRNL-823329}
\preprint{SI-HEP-2021-18}
\title{Goldstone Boson Scattering with a Light Composite Scalar}

\author{T.~Appelquist}
\affiliation{Department of Physics, Sloane Laboratory, Yale University, New Haven, Connecticut 06520, USA}
\author{R.~C.~Brower}
\affiliation{Department of Physics and Center for Computational Science, Boston University, Boston, Massachusetts 02215, USA}
\author{K.~K.~Cushman}
\affiliation{Department of Physics, Sloane Laboratory, Yale University, New Haven, Connecticut 06520, USA}
\author{G.~T.~Fleming}
\affiliation{Department of Physics, Sloane Laboratory, Yale University, New Haven, Connecticut 06520, USA}
\author{A.~Gasbarro}\email{andrew.gasbarro@yale.edu}
\affiliation{AEC Institute for Theoretical Physics, University of Bern, 3012 Bern, CH}
\author{A.~Hasenfratz}
\affiliation{Department of Physics, University of Colorado, Boulder, Colorado 80309, USA}
\author{J.~Ingoldby}\email{ingoldby@ictp.it}
\affiliation{Abdus Salam International Centre for Theoretical Physics, Strada Costiera 11, 34151, Trieste, Italy}
\author{X.~Y.~Jin}
\affiliation{Computational Science Division, Argonne National Laboratory, Argonne, Illinois 60439, USA}
\author{E.~T.~Neil}
\affiliation{Department of Physics, University of Colorado, Boulder, Colorado 80309, USA}
\author{J.~C.~Osborn}
\affiliation{Computational Science Division, Argonne National Laboratory, Argonne, Illinois 60439, USA}
\author{C.~Rebbi}
\affiliation{Department of Physics and Center for Computational Science, Boston University, Boston, Massachusetts 02215, USA}
\author{E.~Rinaldi}
\affiliation{Physics Department, University of Michigan, Ann Arbor, MI 48109, USA}
\affiliation{Theoretical Quantum Physics Laboratory, RIKEN, 2-1 Hirosawa, Wako, Saitama 351-0198, Japan}
\affiliation{Interdisciplinary Theoretical and Mathematical Sciences Program (iTHEMS), RIKEN, 2-1 Hirosawa, Wako, Saitama 351-0198, Japan}
\author{D.~Schaich}
\affiliation{Department of Mathematical Sciences, University of Liverpool, Liverpool L69 7ZL, UK}
\author{P.~Vranas}
\affiliation{Physical and Life Sciences Division, Lawrence Livermore National Laboratory, Livermore, California 94550, USA}
\affiliation{Nuclear Science Division, Lawrence Berkeley National Laboratory, Berkeley, California 94720, USA}
\author{E.~Weinberg}
\affiliation{Department of Physics and Center for Computational Science, Boston University, Boston, Massachusetts 02215, USA}
\affiliation{NVIDIA Corporation, Santa Clara, California 95050, USA}
\author{O.~Witzel}
\affiliation{Center for Particle Physics Siegen (CPPS), Theoretische Physik 1, Naturwissenschaftlich-Technische Fakult\"at, Universit\"at Siegen, 57068 Siegen, Germany}

\collaboration{Lattice Strong Dynamics (LSD) Collaboration}
\noaffiliation


\begin{abstract}
  The appearance of a light composite $0^+$ scalar resonance in nearly conformal gauge--fermion theories motivates further study of the low energy structure of these theories. To this end, we present a nonperturbative lattice calculation of s-wave scattering of Goldstone bosons in the maximal-isospin channel in $\SU(3)$ gauge theory with $N_f=8$ light, degenerate flavors. The scattering phase shift is measured both for different values of the underlying fermion mass and for different values of the scattering momentum. We examine the effect of a light flavor-singlet scalar (reported in earlier studies) on Goldstone boson scattering, employing a dilaton effective field theory (EFT) at the tree level. The EFT gives a good description of the scattering data, insofar as the magnitude of deviations between EFT and lattice data are no larger than the expected size of next-to-leading order corrections in the EFT.
\end{abstract}

\maketitle

\section{\label{sec:intro}Introduction}
The existence of a light composite $0^+$ scalar resonance has been strongly indicated by lattice calculations in a variety of four-dimensional gauge--fermion theories near the lower boundary of the conformal window~\cite{Aoki:2013pca, Aoki:2013zsa, Aoki:2013hla, Aoki:2014oha, Aoki:2016wnc,Athenodorou:2014eua,Fodor:2015vwa,Rinaldi:2015axa,Brower:2015owo,Fodor:2016pls,Hasenfratz:2016gut,DelDebbio:2015byq,Appelquist:2016viq,Appelquist:2018yqe,Gasbarro:2017fmi,Gasbarro:2019kgj,Athenodorou:2017dbf, Fodor:2017nlp,Athenodorou:2021wom,Brower:2019oor}.
Its presence has sparked renewed interest in theories of a composite Higgs boson, e.g.~\cite{BuarqueFranzosi:2018eaj, Appelquist:2020bqj, Witzel:2019jbe, Brower:2019oor, Appelquist:2020xua, Witzel:2020hyr}.

Despite considerable numerical effort to date, little is known from direct lattice calculation about the light scalar resonance beyond its mass. For these studies, work is ongoing to control lattice artifacts~\cite{Fodor:2016pls} and finite-volume effects \cite{Fodor:2020niv} and to push closer to the chiral limit~\cite{Appelquist:2018yqe}.
Lattice computation near the conformal window is generally impeded by the long correlation lengths of the nearly conformal systems, and the disconnected diagrams intrinsic to the evaluation of correlation functions of flavor singlet operators present additional challenges. Beyond the singlet scalar two point function, form factor and scattering amplitudes involving the scalar singlet as an external state are also desirable to study but present similar or greater numerical challenges in practice.

In this work, we instead probe the physics of the light singlet scalar through the lattice computation of pseudo-Nambu--Goldstone boson (PNGB) scattering at low energies.
Chiral symmetry completely constrains PNGB scattering to leading order when the PNGB momenta and mass $M_{\pi}$ are small compared to other scales \cite{Weinberg:1966kf}.
However, the presence of a light $0^+$ scalar of mass $M_\sigma$ leads to additional pole terms in the leading-order expressions for the scattering amplitude whose form cannot be fixed by chiral symmetry~\cite{Lai:1969xr}.
For the lattice ensembles we consider, $M_{\pi} \approx M_\sigma$, indicating that the scalar pole terms may contribute significantly if the coupling between the PNGBs and $0^+$ scalar is not too small.

\begin{table}[h!]
  \centering
  \renewcommand\arraystretch{1.2}  
  \addtolength{\tabcolsep}{1.5 pt} 
  \begin{tabular}{llc|rcrrr}
    \toprule
    Volume            & $am$    & $\tau$ & MDTU  & Sep. & $N_{\text{meas}}$ & $L_{\text{Bin}}$ & $N_{\text{Bin}}$ \\
    \midrule
    \hline
    $24^3 \times 48$  & 0.00889 & 1.0    & 24480 & 40   &  613              & 4                & 153              \\
    \hline
    $32^3 \times 64$  & 0.00889 & 1.0    &  5960 & 40   &  150              & 8                &  18              \\
                      &         &        &  5960 & 40   &  150              & 8                &  18              \\
                      &         &        &  5960 & 40   &  150              & 8                &  18              \\
                      &         &        &  5960 & 40   &  150              & 8                &  18              \\
    \hline
    $24^3 \times 48$  & 0.0075  & 1.0    &  9680 & 40   &  243              & 4                &  60              \\
    \hline
    $32^3 \times 64$  & 0.0075  & 1.0    & 24600 & 40   &  616              & 4                & 154              \\
    \hline
    $48^3 \times 96$  & 0.0075  & 1.0    & 19690 & 10   & 1970              & 8                & 246              \\
                      &         &        & 19580 & 10   & 1959              & 8                & 244              \\
    \hline
    $32^3 \times 64$  & 0.005   & 2.0    &  2520 & 40   &   64              & 1                &  64              \\
                      &         &        &  2520 & 40   &   64              & 1                &  64              \\
                      &         &        &  3160 & 40   &   80              & 1                &  80              \\
    \hline
    $48^3 \times 96$  & 0.005   & 1.0    &  3950 & 10   &   396             & 8                &  49              \\
                      &         &        &  3140 & 10   &   315             & 8                &  39              \\
    \hline
    $48^3 \times 96$  & 0.00222 & 1.0    &  2048 & 16   &  129              & 3                &  43              \\
                      &         &        &  1168 & 16   &   74              & 3                &  24              \\
    \hline
    $64^3 \times 128$ & 0.00125 & 0.5    &  1296 & 36   &   37              & 4                &   9              \\
                      &         &        &  1296 & 36   &   37              & 4                &   9              \\
                      &         &        &  1296 & 36   &   37              & 4                &   9              \\
                      &         &        &  1920 & 40   &   49              & 4                &  12              \\
                      &         &        &  3132 & 36   &   88              & 4                &  22              \\
                      &         &        &  1080 & 40   &   28              & 4                &   7              \\
                      &         &        &  3132 & 36   &   88              & 4                &  22              \\
                      &         &        &  1520 & 40   &   39              & 4                &   9              \\
                      &         &        &  3132 & 36   &   88              & 4                &  22              \\
                      &         &        &   920 & 40   &   24              & 4                &   6              \\
    \bottomrule
  \end{tabular}
  \caption{Summary of statistics for each stream, specifying the lattice volume, the fermion mass $am$, and the trajectory length $\tau$ (in molecular dynamics time units, MDTU) used in the configuration generation via the hybrid Monte Carlo algorithm.  All streams use bare coupling $\beta_F = 4.8$.  For each stream, we perform measurements each ``Sep.'' MDTU, which provides $N_{\text{meas}}$ measurements from the stated number of thermalized MDTU.  We reduce autocorrelations by combining $L_{\text{Bin}}$ measurements into $N_{\text{Bin}}$ `binned' samples, choosing $L_{\text{Bin}}$ by analyzing autocorrelations of the PNGB and $I=2$ correlation functions.}
  \label{tab:ensembles}
\end{table}
In this work we investigate the role of the light scalar in s-wave $\pi \, \pi \to \pi \, \pi$ PNGB scattering at low energies.
We employ the L{\"u}scher finite volume method~\cite{Luscher:1990ux} to extract the scattering phase shift nonperturbatively on the lattice.
The near-conformality of the gauge theory and the observed spectrum reported in previous publications \cite{Appelquist:2016viq,Appelquist:2018yqe} lead us to interpret the lattice results in terms of an effective field theory (EFT) including a light singlet scalar interpreted as an approximate dilaton, along with the PNGBs \cite{Golterman:2016lsd,Appelquist:2017wcg,Appelquist:2017vyy,Appelquist:2019lgk}. Alternative EFT descriptions for the light scalar have been suggested in Refs.~\cite{Appelquist:2018tyt,Hansen:2016fri}.

We focus on a scattering channel, designated as ``maximal isospin'', which contains no fermion-line-disconnected diagrams, allowing us to extract the scattering length to a high statistical precision.
The scalar can then be exchanged in the $t$ and $u$ channels, contributing to the scattering length and effective range. In general, the coupling of the $0^+$ scalar to the PNGBs can depend on the details of explicit chiral symmetry breaking, and may also have dependence on the momenta. In the context of the EFT we employ, it will be determined by only a small number of EFT parameters.

In \secref{sec:luscher}, we detail our numerical lattice calculation of low-energy $\pi\pi$ scattering.
In \secref{sec:dilaton}, we apply the dilaton EFT to the scattering length data, and assess the consistency of the EFT and the ability of the scattering data to constrain the model.
We summarize our results in \secref{sec:conclusion}, and discuss open questions.

\section{\label{sec:luscher}Lattice Computation of Scattering Phase Shift}
\begin{table*}[t]
  \centering
  \renewcommand\arraystretch{1.2}  
  \addtolength{\tabcolsep}{0.5 pt} 
  \begin{tabular}{llllllcllc}
  \toprule
  Volume &  $a m_q$ &     $a M_\pi$ &     $a F_\pi$ & $a M_\sigma$ &      $a^2 k^2$ & $a k \cot\delta$ & $M_\pi^2 / F_\pi^2$ & $k^2 / M_\pi^2$ & $M_\pi / k\cot\delta$ \\
  \midrule
  $24^3 \times 48$ &  0.00889 &   0.22754(21) &  0.051862(77) &    0.279(30) &    0.00193(15) &       -0.585(38) &          19.250(62) &      0.0374(29) &            -0.391(24) \\
  $32^3 \times 64$ &  0.00889 &  0.225501(85) &  0.052449(41) &           -- &   0.000745(43) &       -0.602(35) &          18.485(24) &     0.01466(85) &            -0.376(19) \\
  $24^3 \times 48$ &  0.00750 &   0.20949(28) &  0.047428(94) &           -- &    0.00240(17) &       -0.491(31) &           19.51(10) &      0.0547(40) &            -0.428(25) \\
  $32^3 \times 64$ &  0.00750 &   0.20616(11) &  0.048171(51) &   0.2573(66) &   0.000830(45) &       -0.548(25) &          18.316(36) &      0.0195(11) &            -0.377(17) \\
  $48^3 \times 96$ &  0.00750 &  0.205686(20) &  0.048220(15) &           -- &  0.0002410(29) &      -0.5292(56) &          18.196(11) &    0.005696(68) &           -0.3887(41) \\
  $32^3 \times 64$ &  0.00500 &   0.16725(21) &  0.039196(81) &    0.182(24) &   0.001117(75) &       -0.429(25) &          18.207(78) &      0.0399(27) &            -0.391(21) \\
  $48^3 \times 96$ &  0.00500 &  0.165841(52) &  0.039823(28) &           -- &   0.000264(16) &       -0.489(26) &          17.343(24) &     0.00961(57) &            -0.340(18) \\
  $48^3 \times 96$ &  0.00222 &   0.10972(10) &  0.027104(44) &    0.131(20) &   0.000416(25) &       -0.330(17) &          16.388(60) &      0.0346(22) &            -0.333(17) \\
  $64^3 \times 128$ &  0.00125 &  0.081971(67) &  0.021419(27) &    0.089(32) &   0.000216(10) &       -0.265(11) &          14.646(48) &      0.0321(15) &            -0.309(12) \\
  \bottomrule
  \end{tabular}
  \caption{Physical observables extracted from PNGB and $I=2$ correlation functions on each gauge ensemble, with the addition of the $0^{++}$ mass $M_\sigma$ results computed in Ref.~\cite{Appelquist:2018yqe}. Here, $a$ indicates the lattice spacing.  Error bars represent statistical plus fit systematics determined through the AIC procedure.}
  \label{tab:data}
\end{table*}

We carry out our lattice study of maximal isospin $\pi\pi$ scattering on gauge ensembles for $\SU(3)$ gauge theory with two staggered flavors which become $N_f=8$ dynamical, degenerate light flavors in the continuum limit.
The unrooted staggered fermion action is improved with a single nHYP smearing step~\cite{Hasenfratz:2001hp, Hasenfratz:2007rf, Cheng:2011ic}.
The gauge action includes both fundamental and adjoint plaquette terms.
The phase structure of this lattice action as well as the low lying hadronic spectrum have been presented in Refs.~\cite{Appelquist:2016viq,Appelquist:2018yqe,Gasbarro:2017fmi,Schaich:2015psa}.

In the continuum limit, the four tastes of each staggered flavor become degenerate, and the full $\SU_L(8)\times \SU_R(8) \rightarrow \SU_V(8)$ chiral symmetry breaking pattern is recovered with 63 PNGBs. Any two of the $N_f=8$ fermion flavors can be chosen to transform under an $\SU(2)$ subgroup of this $\SU(8)$, with the remaining 6 transforming as singlets.
For $2 \rightarrow 2$ PNGB scattering in the maximal $I=2$ isospin channel of this $\SU(2)$ subgroup, there is therefore no mixing with two particle states involving the other 6 flavors. Equivalently, two PNGBs in an $I=2$ state also belong to an irreducible representation of the full $\SU(8)$. This property makes a complete coupled channel analysis unnecessary.

To better understand how lattice artifacts affect the global symmetries, consider the subgroup $\SU_V(2)\times \SU_V(4) \subset \SU_V(8)$.
This $\SU_V(2)$ rotates between the two staggered flavors and is exact even at finite lattice spacing, and $\SU_V(4)$ is the continuum symmetry for a single staggered flavor.
The four tastes of each staggered flavor lead to 16 tastes of pseudoscalar bilinears, an $\SU(4)$ adjoint $\mathbf{15}$ and singlet $\textbf{1}$.
Each of these 16 pseudoscalar tastes may transform in the adjoint $\mathbf{3}$ or singlet $\mathbf{1}$ representation of $\SU_V(2)$, for a total of 64 pseudoscalars corresponding to the adjoint $\mathbf{63}$ and singlet $\mathbf{1}$ of $\SU_V(8)$.
The 16 tastes are designated by the spin-taste structure $\gamma_5 \bigotimes \xi_F$ \cite{Daniel:1987aa}.

At finite lattice spacing, the $\SU_V(4)$ taste symmetry is explicitly broken at $\cO(a^2)$ to $\SO(4)$ and at $\cO(a^2 p^2)$ to $\text{SW}_4$, the discrete lattice spin-taste group \cite{Kilcup:1986dg,Sharpe:2004is}.
The $\gamma_5 \bigotimes \xi_5$ non-singlet pseudoscalars become the only three exact PNGBs in the chiral limit, with the remaining sixty having masses proportional to the lattice spacing.
The spectral study has revealed that the taste splitting of the 63-plet masses are on the order of 20--30\% \cite{Appelquist:2018yqe}.
Lattice artifacts generate irrelevant four-fermion operators which mix the $\xi_F$ tastes, so that maximal isospin $2\rightarrow 2$ scatting can involve coupling between two particle states of other tastes at finite lattice spacing.
However, such mixing should be highly suppressed, particularly for the scattering of pseudoscalars in an $\SO(4)$ irrep at low momentum.
Therefore, we will assume it sufficient to study a single (diagonal) taste channel.

We consider the $2\rightarrow 2$ s-wave scattering of pseudoscalars with spin-taste structure $\gamma_5 \bigotimes \xi_5$ in the maximal isospin $I=2$ channel of $\SU_V(2)$.
Denoting the staggered flavors as $(\chi_1,\chi_2)$, this spin-taste structure corresponds to the distance-zero interpolating operator $\pi^+(x) = \bar\chi_2(x) \epsilon(x)\chi_1(x)$, with staggered phase $\epsilon(x) = (-1)^{x+y+z+t}$.
We probe the $I=2$ s-wave scattering channel with the interpolating operator constructed from two PNGBs at rest.
\begin{equation}
\cO_{I=2}(t) = \pi^+(t)\pi^+(t+1)
\end{equation}
where the individual PNGB operators are projected onto zero spatial momentum
\begin{equation}
\pi^+(t) = \sum_{\vec{x}} \bar\chi_2(x) \epsilon(x)\chi_1(x).
\end{equation}
We separate the operators by one time slice to avoid the projection onto unwanted states from Fiertz rearrangement identities \cite{Kuramashi:1993ka,Fukugita:1994ve,Fu:2013ffa}.
The $I=2$ two point correlation function is then given by
\begin{flalign}
  & C_{I=2}(t,t_0) = \la \cO_{I=2}(t) \cO_{I=2}(t_0)^\dagger \ra \\
               & =  \sum_{\vec{x}_1, \cdots, \vec{x}_4} \la \pi^+(t_4,\vec{x}_4) \pi^+(t_3,\vec{x}_3) \pi^+(t_2,\vec{x}_2)^\dagger \pi^+(t_1,\vec{x}_1)^\dagger \ra \nonumber
\end{flalign}
with $t_1 = t_0$, $t_2 = t_0+1$, $t_3 = t$, and $t_4 = t+1$.
Two types of Wick contractions contribute to the correlation function, the so called ``direct'' (D) and ``crossed'' (C) channels.
\begin{align}
  C_{I=2}(t,t_0) & = C_D(13;24) + C_D(14;23) \cr
                 & \quad - C_C(1324) - C_C(1423) , \\
  C_D(ik;jl) & = \sum_{\substack{\vec{x}_i,\vec{x}_j,\\\vec{x}_k,\vec{x}_l}}\text{Tr}\Big[G_{x_k x_i}^\dagger G_{x_k x_i}\Big]\Tr{G_{x_l x_j}^\dagger G_{x_l x_j}},\;\;\;\; \label{eq:directpoint} \\
  C_C(ijkl)  & = \sum_{\substack{\vec{x}_i,\vec{x}_j,\\\vec{x}_k,\vec{x}_l}}\Tr{G_{x_k x_i}G_{x_k x_j}^\dagger G_{x_l x_j} G_{x_l x_i}^\dagger },    \label{eq:crossedpoint}
\end{align}
where $\Tr{\dots}$ denotes the color trace and $G_{xy}$ is a point propagator from $y$ to $x$.
The $C_D$ and $C_C$ correlators may instead be constructed from Green's functions computed on wall sources \cite{Kuramashi:1993ka,Fukugita:1994ve} satisfying
\begin{equation}
  \sum_{x'} D_{x,x'} G_{x',t_0}^W = \sum_{\vec{y}} \delta_{x,(\vec{y},t_0)}.
\end{equation}
By inverting on a wall source at $t_i=t_0$ and $t_j=t_0+1$, we construct the Wick diagrams for the $I=2$ correlation function with only $2 N_c$ inversions per configuration:
\begin{flalign}
C_D^W(ik;jl) & = \sum_{\vec{x}_k,\vec{x}_l} \Tr{G_{x_k, t_i}^{W \;\dagger} G^W_{x_k, t_i}}\Tr{G_{x_l, t_j}^{W \;\dagger} G^W_{x_l, t_j}} , \\
C_C^W(ijkl)  & = \sum_{\vec{x}_k,\vec{x}_l} \Tr{ G^W_{x_k, t_i}G_{x_k, t_j}^{W \; \dagger} G^W_{x_l, t_j} G_{x_l, t_i}^{W \; \dagger} } ,
\end{flalign}
which are equivalent to Eqs.~(\ref{eq:directpoint}) and (\ref{eq:crossedpoint}) up to terms with open quark lines which produce gauge-variant noise.
We choose not to gauge fix, allowing the gauge variant terms to average out in the ensemble average rather than being removed by explicit gauge fixing.
This is sometimes referred to as the ``moving wall'' method \cite{Fu:2013ffa}.
From the same wall sources, we also construct the two point correlation function of $\pi^+$ at rest and extract the PNGB mass and decay constant.

In this work we restrict our attention to low momentum scattering, where the scattering phase shift may be expanded in powers of $k^2/M_\pi^2$.
\begin{equation}
k\cot\delta^{I}(k) = \frac{1}{a^I} + \frac{1}{2} r^I M_\pi^2 \left( \frac{k^2}{M_\pi^2} \right) + \cO\left( \frac{k^4}{M_\pi^4} \right),\label{eq:effrexp}
\end{equation}
where $a^I$ and $r^I$ are the s-wave scattering length and effective range respectively in the isospin $I$ channel. We have suppressed angular momentum indices on these quantities.

The gauge ensembles used in this study are detailed in Table~\ref{tab:ensembles}.
A multiexponential fitting procedure is applied to the $I=2$ correlation function as well as the $\pi^+$ two point function.
On each gauge ensemble and for each correlation function, we apply a consistent fitting procedure and statistical analysis.
Energy levels are extracted with a three state $\cosh$ fit using the full covariance matrix between time slices.
In the case of the $I=2$ correlation function, an extra constant is added to the fit function to account for a wrap around effect where each of the PNGBs propagate in opposite directions.
The second excited state, being the highest fit state in the correlation function, is assumed to be excited state contaminated and is discarded.
In principle the first $I=2$ excited state can be used to probe larger scattering momenta at a fixed quark mass.
However, we have found that the first excited state energies on our ensembles correspond to scattering momenta outside of the radius of convergence of the effective range expansion Eq.~(\ref{eq:effrexp}).
Therefore in this study we focus our attention entirely on the ground state $I=2$ energy level on each ensemble.

To completely control the fit range systematics, we apply an extension of the Akaike information criterion (AIC)~\cite{Akaike:1974AIC} studied recently by Jay and Neil to average over fits to correlation functions in lattice QCD~\cite{Jay:2020jkz}.
Each fit is assigned a probability $\text{Pr}(M|D)$ as
\begin{flalign}
&-2\ln\left( \text{Pr}(M|D) \right) = \nonumber\\
& \;\;\;\;-2\ln\left( \text{Pr}(M) \right) + \chi^2_{\text{min}} + 2k + 2N_\text{cut},
\end{flalign}
where $\text{Pr}(M)$ is the prior given to each fit, which we take to be uniform, $\chi^2_{\text{min}}$ is the minimum $\chi^2$ which is $-2\times\text{(max log likelihood)}$ for each fit, $k$ is the number of model parameters used in the fit, and $N_\text{cut}$ is the number of data points cut out by the fit window.
The extended AIC probability is then used to average over the best fit values of the parameters in each fit, as well as to average the bootstrap error bars.
In our fits, we have a good signal to noise ratio all the way across the lattice, allowing us to always take $t_\text{max} = T/2$.
Therefore, we need only control the $t_\text{min}$ systematic through this procedure.
In all cases, we find that the AIC averaged result is in good agreement with the value extracted from picking a single fit where the excited state contamination has plateaued with $t_\text{min}$, though the AIC average generally assigns a slightly larger overall error bar than choosing a single fit.

The scattering phase shift is computed via the L{\"u}scher procedure from the ground state energy of the $I=2$ correlation function.
For two identical scalars in an s-wave, the L{\"u}scher finite volume quantization condition \cite{Luscher:1990ux} is given by
\begin{eqnarray}
&k\cot\delta(k) = \frac{2\pi}{L} \pi^{-3/2} Z_{00}\left(1,\frac{k^2L^2}{4\pi^2} \right) , \\
&k^2 = \frac{1}{4} E_{\pi\pi}^2 - M_\pi^2 \; , \; Z_{00}(1,q^2) = \frac{1}{\sqrt{4\pi}} \sum\limits_{\vec{n}\in\mathbb{Z}^3} \frac{1}{\vec{n}^2 - q^2}.
\end{eqnarray}
Each two-PNGB energy level extracted from the $I=2$ correlation function yields a corresponding determination of the scattering phase shift at scattering momentum $k^2$.
The scattering observables $k^2$ and $k\cot\delta$ depend systematically on the fit range choice for both the $I=2$ and PNGB correlation functions.
This systematic is controlled by multiplying the AIC distributions to form a joint distribution in $(t_{\text{min},\pi},t_{\text{min},I=2})$ and averaging against this distribution.

The results of the multiexponential fitting and scattering analysis on each ensemble are shown in Table~\ref{tab:data}, where errors reflect statistical plus fit systematic errors processed through the AIC procedure.
The values of the PNGB mass and decay constant reported here are from the new analysis with wall sources as described above. We use the normalization of the PNGB decay constant corresponding to the QCD value $F_\pi = 92.2(1)$ MeV as in Ref.~\cite{Appelquist:2018yqe}. 
The results are consistent within statistical errors compared to Ref.~\cite{Appelquist:2018yqe}, but errors have shrunk considerably due a combination of the fit range systematic control through the AIC procedure, higher availability of gauge statistics for certain ensembles, and perhaps a statistics gain through the use of wall sources. We have restated the values of the singlet scalar mass computed in~\cite{Appelquist:2018yqe} for convenience because we will make use of it in Section.~\ref{sec:dilaton}. In addition to the basic quantities in lattice units, we have included values for certain ratios that will be useful in Section~\ref{sec:dilaton}.

\begin{figure}[t!]
  \centering
  \includegraphics[width=0.9\columnwidth]{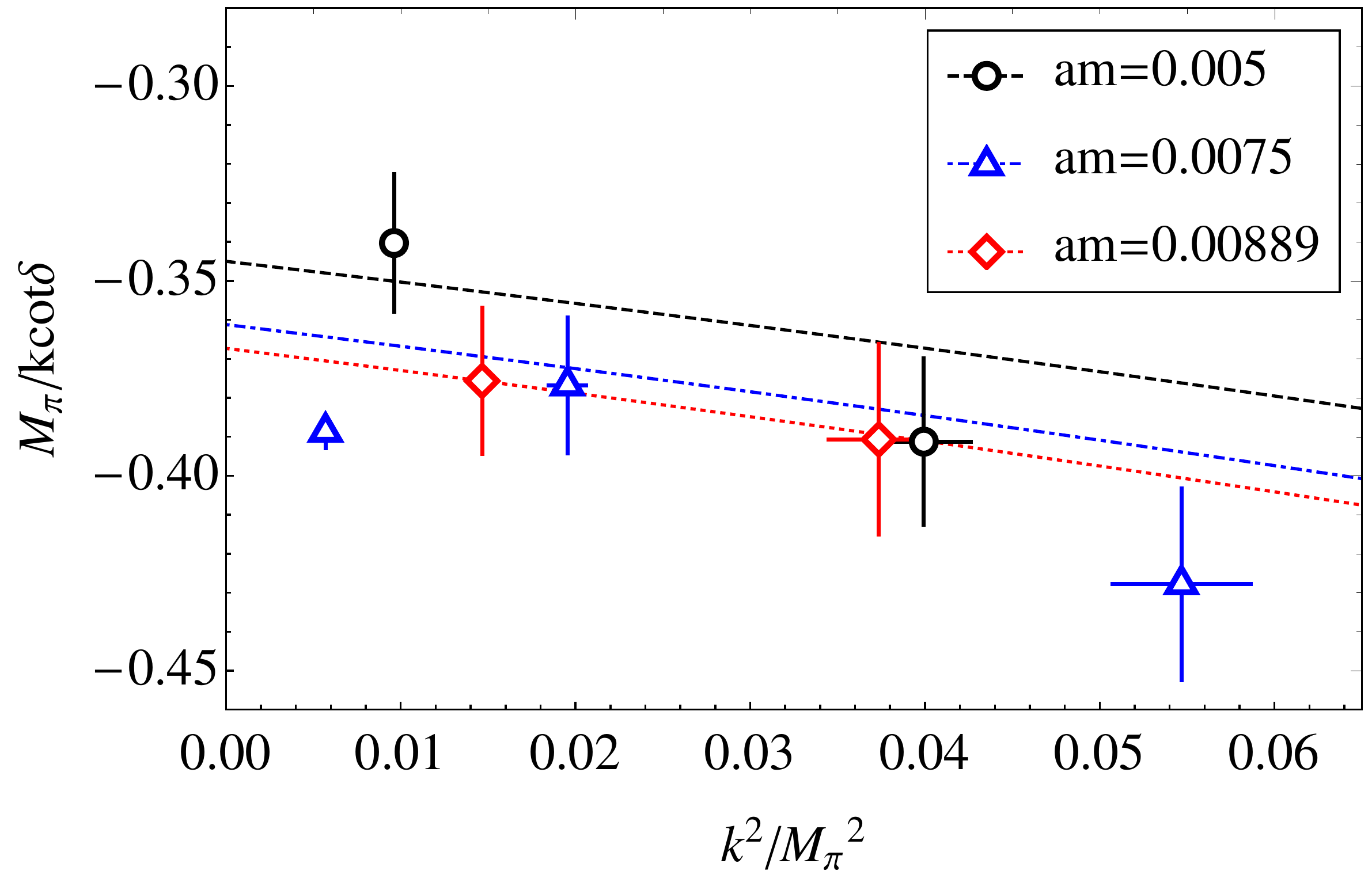}
  \caption{Plot showing the variation of the scattering phase shift with scattering momentum for the three largest fermion masses. The points indicate the lattice data, given in Table~\ref{tab:data}, whereas the lines represent tree level dilaton EFT predictions, as explained in Section \ref{sec:dilaton}.\label{fig:k2dep}  }
\end{figure}

The scattering length and effective range are in principle independent observables, and are sensitive to a different set of low energy parameters in an EFT. We assess the significance of the effective range contribution to the scattering phase shift in \fig{fig:k2dep} by plotting $M_\pi / k\cot\delta$ against the effective range expansion parameter for the gauge ensembles for which we have two volumes with the same $a m$.
For comparison we show the leading order dilaton EFT prediction with the dotted and dashed lines.
These curves have been determined by fixing the dilaton EFT parameters through a global fitting procedure which we explain in detail below.
While the data points for the same $am$ tend to lie within error bars, data shows a clear downward trend in correspondence with the dilaton EFT prediction, indicating a non-negligible sensitivity to the effective range term.
The leftmost point with blue triangular marker corresponds to the ensemble with $a m = 0.0075$ and volume $48^3\times 96$ where we have very high statistics.
This point is in clear tension with the tree level dilaton EFT prediction.
The overall validity of the tree level dilaton EFT description of the scattering data is discussed further in Section~\ref{sec:dilaton}.

\subsection{Accounting for Finite--Volume Effects}
\label{sec:FV}

To compare the lattice data of Table~\ref{tab:data} with the predictions of a continuum EFT, we must first ensure that finite-volume effects in all lattice determined quantities are small relative to their quoted uncertainties. We have data taken for many quantities at multiple lattice volumes using the same fermion mass (overlapping volumes), enabling estimation of these effects.

It can be seen in Table~\ref{tab:data} that central values for $aM_\pi$ and $aF_\pi$ taken using the same fermion mass change by an amount greater than their statistical uncertainties between lattice volumes. We therefore perform a dedicated infinite volume extrapolation of the PNGB masses and decay constants in this section, and use this extrapolated data in the fit to dilaton EFT in section~\ref{sec:dilaton}.

We do not have data for $aM_\sigma$ at multiple lattice volumes, so the volume dependence cannot be inferred directly from data, and we will not perform a dedicated infinite--volume extrapolation for this quantity. We will attribute the volume dependence of the scattering phase shift in Table~\ref{tab:data} entirely to the scattering momentum $k^2$ dependence coming from the effective range, as discussed above and illustrated in Fig~\ref{fig:k2dep}.

To extrapolate $aM_\pi$ and $aF_\pi$ to the infinite volume limit, we use the following model
\begin{align}
  M_\pi(L) & = M_\pi\left(1+\alpha\frac{M^2_\pi}{F^2_\pi}\frac{e^{-M_\pi L}}{(M_\pi L)^{3/2}}\right),\label{eq:mpil}\\
  F_\pi(L) & = F_\pi\left(1-\beta\frac{M^2_\pi}{F^2_\pi}\frac{e^{-M_\pi L}}{(M_\pi L)^{3/2}}\right).\label{eq:fpil}
\end{align}
In these formulae, $M_\pi$ and $F_\pi$ represent the extrapolated mass and decay constant. The parameters $\alpha$ and $\beta$ are extracted from a fit to lattice data, and crucially, we assume that they have no dependence on the fermion mass.
Expressions similar to Eqs.~(\ref{eq:mpil}) and (\ref{eq:fpil}) have been derived earlier in the literature~\cite{Gasser:1986vb,Leutwyler:1987ak,Gasser:1987zq,Luscher:1985dn,Colangelo:2003hf,Colangelo:2004xr}, assuming that the PNGBs are well described by chiral perturbation theory.

We perform a least-squares method fit of Eqs.~(\ref{eq:mpil}) and (\ref{eq:fpil}) to all the data we have in Table~\ref{tab:data} for the PNGB masses and decay constants on overlapping volumes (14 points total). Correlations between the lattice measurements are accounted for in this fit. We find that $\chi^2_\text{min}/N_\text{dof}=2.05$, indicating slight tension between the data and this crude model. The central values and standard errors for $\alpha$ and $\beta$ indicated by the fit are
\begin{equation}
\alpha = 1.530(78),\qquad \beta = 1.62(12).\label{eq:ab}
\end{equation} 

\begin{figure}[t!]
  \centering
  \includegraphics[width=\columnwidth]{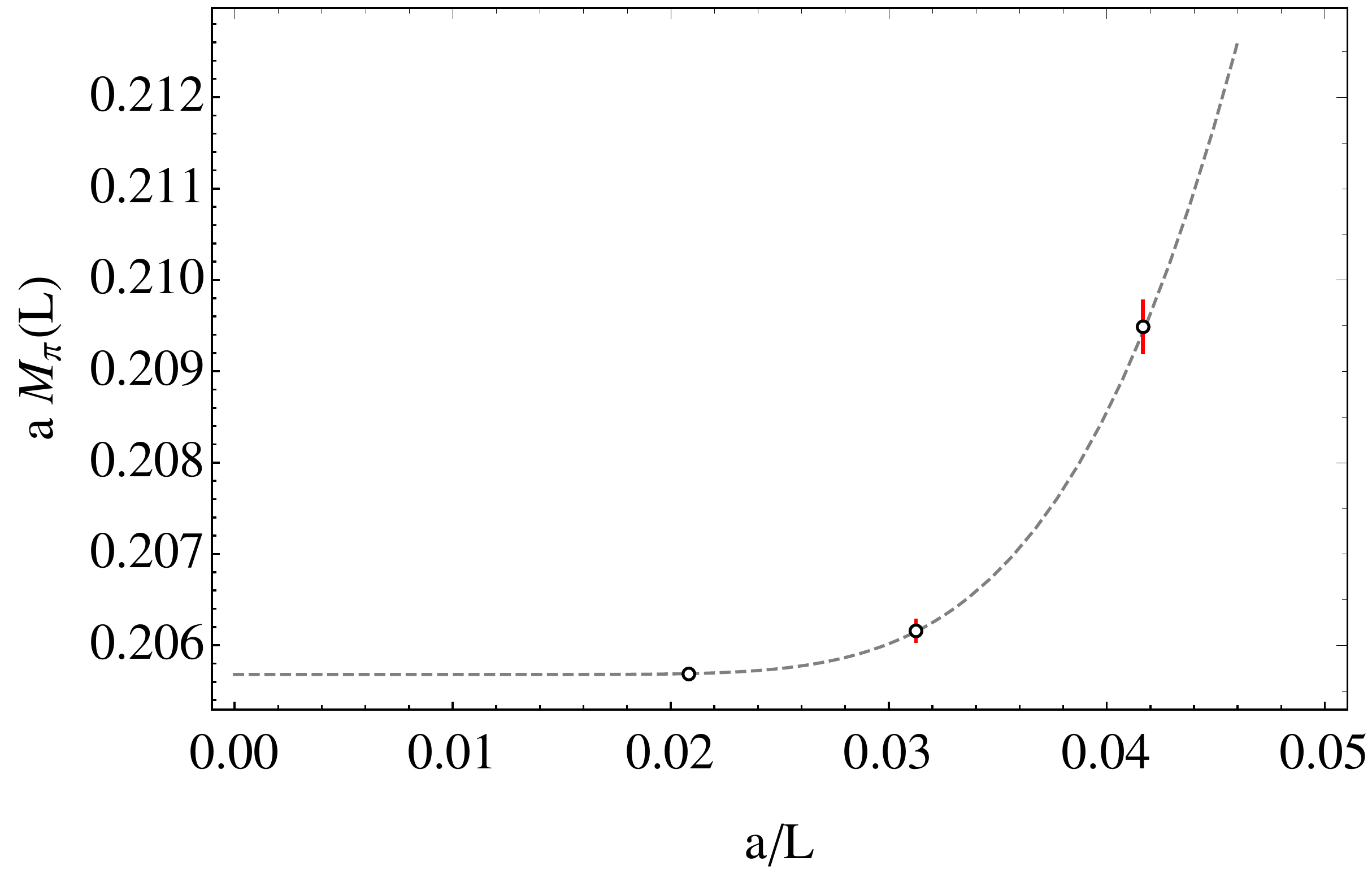}
  \caption{Dependence of the PNGB mass $aM_\pi$ on the inverse of the spatial extent of the lattice $L$ for the fermion mass $am=0.0075$. The gray line shows the $L$--dependence predicted by the model in Eq.~(\ref{eq:mpil}), using the best fit values for the parameters $\alpha$ and $\beta$ indicated in Eq.~(\ref{eq:ab}). For the range of lattice volumes for which we have data, the finite--volume correction to $aM_\pi$ is clearly significant. \label{fig:mpiLdep}  }
\end{figure}

\begin{figure}[t!]
  \centering
  \includegraphics[width=\columnwidth]{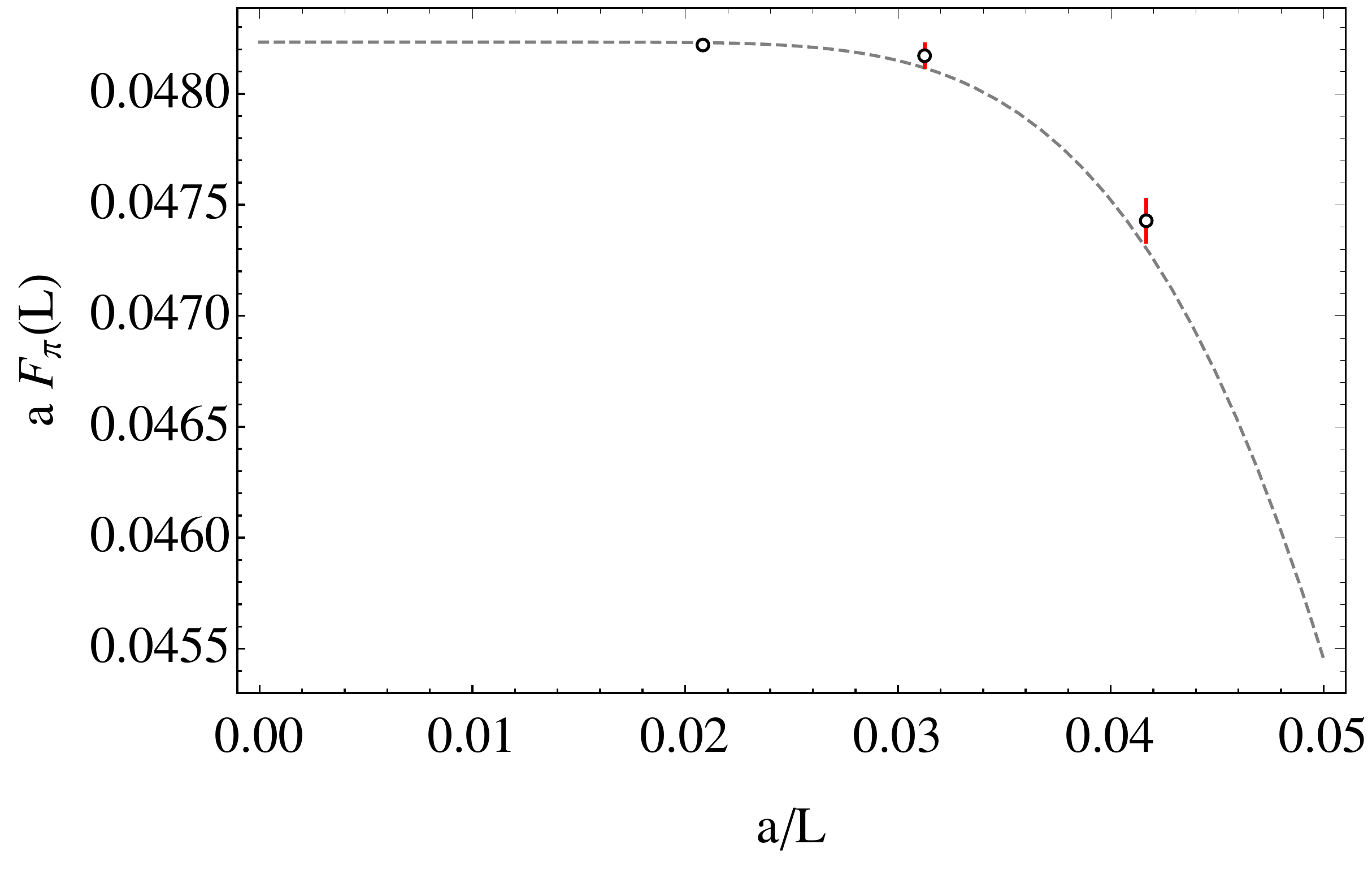}
  \caption{Dependence of the PNGB mass $aF_\pi$ on the inverse of the spatial extent of the lattice $L$ for the fermion mass $am=0.0075$. The gray line shows the $L$--dependence predicted by the model in Eq.~(\ref{eq:fpil}). \label{fig:fpiLdep}  }
\end{figure}

In Figs~\ref{fig:mpiLdep} and \ref{fig:fpiLdep}, lattice data at $am=0.0075$ are plotted on top of the curves given by Eqs.~(\ref{eq:mpil}) and (\ref{eq:fpil}), using parameter values taken from the fit. The plots confirm that the variation in the PNGB mass and decay constant with lattice size is comfortably larger than the statistical uncertainties, and that the model describes this variation reasonably well.

We use this model to estimate the infinite volume extrapolated values of $aM_\pi$ and $aF_\pi$ at all of the remaining fermion masses, for which data at only one lattice volume is available. The results of all the extrapolations are presented in Table~\ref{tab:extra}. This extrapolated data will be compared with EFT predictions in the following section.

\begin{table}[h!]
  \centering
  \renewcommand\arraystretch{1.2}  
  \addtolength{\tabcolsep}{4.0 pt} 
  \begin{tabular}{l l l}
    \toprule
    $am$ & $aM_\pi$ & $aF_\pi$ \\
    \midrule
    0.00889 & 0.22525(27)  & 0.052491(59) \\
    0.0075  & 0.205680(20) & 0.048233(20) \\
    0.005   & 0.16575(10)  & 0.03982(28) \\
    0.00222 & 0.1085(12)   & 0.02742(32) \\
    0.00125 & 0.08115(82)  & 0.02165(23) \\
    \bottomrule
  \end{tabular}
  \caption{Result of the extrapolation $aM_\pi$ and $aF_\pi$ to the infinite volume limit. The central values and uncertainties for these quantities will be used in the fit to Dilaton EFT in section~\ref{sec:dilaton}. Details on how these extrapolations were performed are provided in the text.}
  \label{tab:extra}
\end{table}

We have estimated the total uncertainty in the extrapolated quantities by adding in quadrature their statistical uncertainties (shown in Table~\ref{tab:data}) to the shift in central value due to the infinite volume extrapolation. This shift is given by the difference between that quantity at the largest available lattice volume in Table~\ref{tab:data} and the corresponding central value of that quantity in Table~\ref{tab:extra}. We assigned uncertainties in this conservative way, because the model we are using for these extrapolations is only approximate. These total uncertainties are presented in Table~\ref{tab:extra}. 

As the uncertainties in Table~\ref{tab:extra} arise from systematic finite volume effects to a great extent, which we only estimate using an approximate model, we should not assume that correlations between these extrapolated quantities can be properly obtained from the statistical correlations between the lattice measurements in Table~\ref{tab:data}. We therefore regard all quantities in Table~\ref{tab:extra} as completely uncorrelated.

\section{Dilaton EFT}
\label{sec:dilaton}

The EFT which we employ describes the interaction of a single scalar dilaton with the $N_f^2-1$ PNGBs associated with the breaking of the approximate global symmetry group $G=\SU(N_f)_L\times\SU(N_f)_R$ to the diagonal subgroup $\SU(N_f)_V$.
It is employed to compute the low energy properties of an underlying gauge theory capable of producing a light composite scalar along with the composite PNGBs.
In terms of the PNGB fields $\pi$ and an additional real scalar field $\chi$, the Lagrangian of the EFT is given below. See \cite{Appelquist:2019lgk} and references therein.
\begin{equation}
  \cL = \frac{1}{2}\partial_{\mu}\chi\partial^{\mu}\chi \, + \cL_{\pi} \, + \, \cL_M \, - \, V_{\Delta}(\chi) \, ,
  \label{Eq:L}
\end{equation}
where
\begin{equation}
  \cL_{\pi} = \frac{f_{\pi}^2}{4}\left(\frac{\chi}{f_d}\right)^2 \, \Tr{\partial_{\mu}\Sigma(\partial^{\mu}\Sigma)^{\dagger}} \, .
  \label{Eq:Lpi}
\end{equation}
The matrix field $\Sigma=\exp\left[2i\pi^aT^a/f_{\pi}\right]$ transforms as $\Sigma \to \U_L\Sigma\U_R^{\dagger}$ under the action of unitary transformations $\U_{L,R}\in \SU(N_f)_{L,R}$.
It also satisfies the non-linear constraints $\Sigma\Sigma^{\dagger}=\mathbb{1}_{N_f}$.
The $T^a$ are the generators of $\SU(8)$, normalized so that $\Tr{T^a T^b} = \frac{1}{2} \delta^{ab}$.

The explicit breaking of the global internal symmetry, necessary for the implementation of lattice computations, is described by the term
\begin{equation}
  \cL_M = \frac{m_{\pi}^2 f_{\pi}^2}{4} \left(\frac{\chi}{f_d}\right)^y \, \Tr{\Sigma + \Sigma^{\dagger}} \, ,
  \label{Eq:LM}
\end{equation}
where $m_{\pi}^2 \equiv 2B_{\pi} m$ vanishes when the fermion mass $m$ of the underlying theory is set to zero.
The parameter $y$ can be interpreted as the scaling dimension of the chiral condensate of the underlying gauge theory at strong coupling~\cite{Leung:1989hw}.

The dilaton potential $V_{\Delta}(\chi)$, employed in \refcite{Appelquist:2019lgk} and discussed earlier in Refs.~\cite{Chacko:2012sy,Cata:2018wzl}, is given by
\begin{equation}
  V_{\Delta}(\chi) = \frac{m_d^2\chi^4}{4(4-\Delta)f_d^2}\left[1-\frac{4}{\Delta}\left(\frac{\chi}{f_d}\right)^{\Delta-4} \right] \, .
  \label{Eq:potential}
\end{equation}
It contains two contributions.
One is a scale-invariant term ($\propto \chi^4$) representing the corresponding operators in the underlying gauge theory.
The other ($\propto \chi^{\Delta}$) captures the leading-order effect of the scale deformation in the underlying theory.
The potential has a minimum at $\chi = f_d > 0$, with mass $m_d$ for the dilaton.

In \refcite{Appelquist:2019lgk}, this EFT was used to fit lattice data for the $\SU(3)$ gauge theory with $N_f=8$ Dirac fundamental fermions.
A key feature of the EFT is that the dilaton mass (the explicit breaking of scale symmetry) can be tuned as small as necessary with $f_d$ held fixed.
It has been suggested in Refs.~\cite{Appelquist:2010gy, Golterman:2016lsd,Golterman:2020tdq,Golterman:2020utm} that the explicit breaking in the underlying theory, and also in the EFT as a consequence, can be made arbitrarily small by tuning the number of flavors $N_f$ arbitrarily close to the critical value $N_f^c$ at which confinement gives way to IR conformality.
The form of $V_{\Delta}(\chi)$ interpolates among several specific forms found in the literature.
The choice $\Delta = 2$ gives the Higgs potential of the standard model, while the choice $\Delta \rightarrow 4$ corresponds to a nearly marginal deformation of scale symmetry~\cite{Goldberger:2007zk}.

The mass deformation in \eq{Eq:LM} contributes, in the vacuum $\vev{\Sigma} = \mathbb{1}$, an additive term to $V_{\Delta}$.
The entire potential is
\begin{equation}
  W(\chi) = V_{\Delta}(\chi) \, - \, \frac{N_f m_{\pi}^2f_{\pi}^2}{2} \left(\frac{\chi}{f_d}\right)^y \, ,
  \label{Eq:Wpotential}
\end{equation}
leading to a new minimum for $\chi$ which determines its vacuum value $\vev{\chi} = F_d > f_d$.
Minimizing the potential leads to the transcendental equation for $F_d$ shown below
\begin{equation}
\frac{F_d^{4-y}}{(4-\Delta)f^{4-y}_d}\left[1-\left(\frac{f_d}{F_d}\right)^{4-\Delta}\right]=\frac{yN_ff^2_\pi m^2_\pi}{2f^2_dm^2_d},\label{eq:minimum}
\end{equation}
which can be solved numerically to obtain $F_d$ for a given choice of the fermion mass and EFT parameters.

By employing the value $\vev{\chi} = F_d$ in Eqs.~(\ref{Eq:Lpi}) and (\ref{Eq:LM}), and properly normalising the PNGB kinetic term, one has simple scaling relations for the PNGB decay constant $F_{\pi}$ and PNGB mass $M_{\pi}$:
\begin{align}
  \frac{F_{\pi}^2}{f_{\pi}^2} & = \frac{F_{d}^2}{f_{d}^2} \, , \label{Eq:scaling} \\
  \frac{M_{\pi}^2}{m_{\pi}^2} & = \left( \frac{F_{d}^2}{f_{d}^2}\right)^{\frac{y}{2}-1} \, . \label{Eq:Mpiscaling}
\end{align}

There is also a new curvature at the potential minimum, determining the dilaton mass $M_d$.
In terms of $F_d/f_d$:
\begin{equation}
  \label{eq:Mdscaling}
  \frac{M^2_d}{F^2_d} = \frac{m^2_d}{(4-\Delta)f^2_d}\left(4-y+(y-\Delta)\left(\frac{f_d}{F_d}\right)^{4-\Delta}\right).
\end{equation}

\subsection{PNGB Scattering}
\begin{figure}
    \includegraphics[width=\columnwidth]{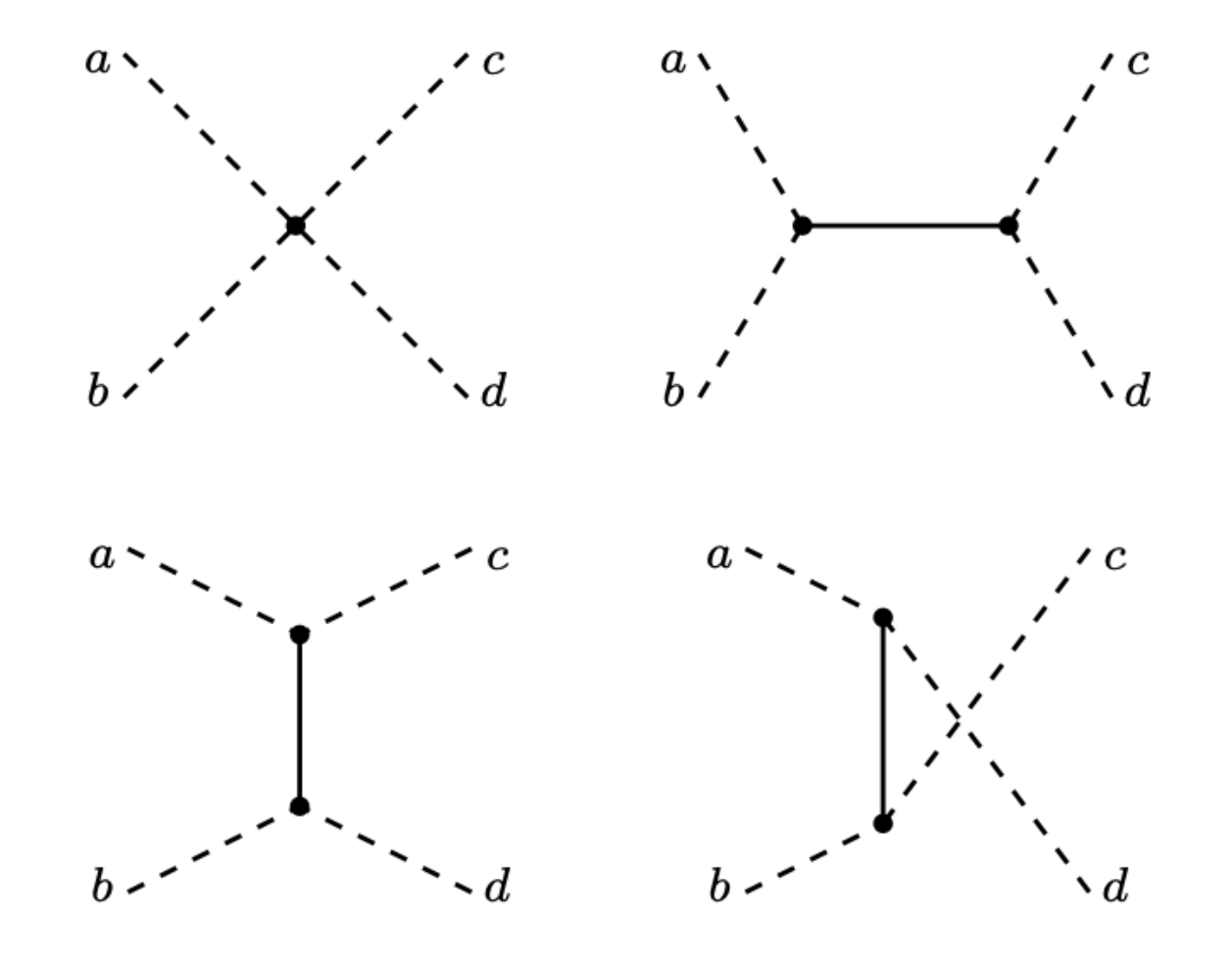}
    \caption{Feynman diagrams in dilaton EFT which appear in the $\pi$--$\pi$ scattering amplitude at tree level. The dashed lines represent PNGBs and the solid lines represent the dilaton.}
    \label{Fig:Feynman}
\end{figure}

The Feynman diagrams that contribute to the scattering process $\pi^a \, \pi^b \to \pi^c \,\pi^d $ at tree level are shown in \fig{Fig:Feynman}.
Here $(abcd) \equiv \Tr{T^a T^b T^c T^d}$, all four momenta are directed inward, and $s \equiv (p_a + p_b)^2$, $ t\equiv (p_a + p_c)^2$ and $u \equiv (p_a + p_d)^2$ are the Mandelstam variables.

The dimensionless scattering amplitude is then given at tree level by
\begin{widetext}
\begin{align}
  i\cM^{ab,cd}(s,t,u) = & \frac{2i}{3F_{\pi}^2}\left\{[(abcd)+(adcb)](6M_{\pi}^2 - 3t) + [(acdb)+(abdc)](6M_{\pi}^2 - 3u) + [(adbc)+(acbd)](6M_{\pi}^2 - 3s)\right\} \nonumber \\
  & -i\frac{\delta^{ab}\delta^{cd}}{F_d^2}\frac{(s+(y-2)M_{\pi}^2)^2}{s-M_d^2}-i\frac{\delta^{ac}\delta^{bd}}{F_d^2}\frac{(t+(y-2)M_{\pi}^2)^2}{t-M_d^2}-i\frac{\delta^{ad}\delta^{bc}}{F_d^2}\frac{(u+(y-2)M_{\pi}^2)^2}{u-M_d^2}.\label{eq:fullamp}
\end{align}
\end{widetext}
We examine a scattering process in which the PNGBs belong to the same $\SU(2)$ triplet, that is, they are each bound states of a single $\SU(2)$ fermion doublet.
When $N_f > 2$, for example $N_f = 8$ in our case, one could imagine that the chosen fermion doublet carries electroweak quantum numbers and the others are electroweak singlets.
The triplet of $\SU(2)$ PNGBs can scatter in three independent isospin channels $I = 0$, $1$ and $2$.
We focus here on the $I= 2$ channel since, as noted in \secref{sec:luscher}, the lattice computation in the underlying gauge theory then contains no fermion-line-disconnected diagrams.
The scattering amplitude in this channel takes the form
\begin{align}
  \cM^{I=2}(s,t,u) = & \frac{2M_{\pi}^2-s}{F_{\pi}^2}-\frac{1}{F_d^2}\frac{(t+(y-2)M_{\pi}^2)^2}{t-M_d^2} \nonumber \\
  & -\frac{1}{F_d^2}\frac{(u+(y-2)M_{\pi}^2)^2}{u-M_d^2}. \label{eq:i=2dilaton}
\end{align}
The first term in this expression comes from the chiral-Lagrangian four-point interaction while the next two (pole) terms arise from exchange of the dilaton scalar in the $t$ and $u$ channels.

For our purposes, we develop $\cM^{I=2}(s,t,u)$ in a partial wave expansion:
\begin{align}
  \cM^I(s,t,u) = 32\pi\sum_{\ell=0}^{\infty}(2\ell+1)P_{\ell}(\cos\theta)t^I_{\ell}(s), \\
  t^I_{\ell}(s) = \frac{1}{64\pi}\int_{-1}^{+1}d(\cos\theta)P_{\ell}(\cos\theta) \cM^I(s,t,u). \label{eq:partialwaves}
\end{align}
For the $\ell=0$ partial wave component, this gives
\begin{align}
  t^{2}_0(s)= &\frac{2M_{\pi}^2-s}{32\pi F_{\pi}^2}+\frac{2k^2-M_d^2-2M_{\pi}^2(y-2)}{16\pi F_d^2} \nonumber \\
  & +\frac{\left[M_d^2+(y-2)M_{\pi}^2\right]^2}{64\pi F_d^2 k^2}\log\left(\frac{4k^2+M_d^2}{M_d^2}\right).\label{eq:i2l0amp}
\end{align}
with $s = 4(k^2 + M_{\pi}^2)$ and $k = |\vec{k}|$ is the PNGB spatial momentum in the center of mass frame.

For any partial wave and isospin channel, the tree-level amplitude $t_{\ell}^{I}(s)$ can be expressed in terms of the scattering phase shift $\delta_{\ell}^{I}$ as
\begin{align}
t^{I}_\ell(s) = \frac{\sqrt{M^2_\pi +k^2}}{k\cot\delta^{I}_\ell-ik},\label{eq:phaseshift}
\end{align}
where $k^{2l+1} \cot\delta_{\ell}$ has a power-series expansion in $k^2$ valid for $k^2 \ll M_{\pi}^2$.
For the case $\ell = 0$, this expansion takes the form shown in \eq{eq:effrexp}.

For the dilaton EFT, we then have
\begin{align}
 M_\pi a^{I=2} = -\frac{M^2_\pi}{16\pi F^2_\pi}\left(1-(y-2)^2\frac{f^2_\pi}{f^2_d}\frac{M^2_\pi}{M^2_d}\right),
\label{eq:dilatonslength}
\end{align}
and
\begin{align}
M_\pi r^{I=2} = & -\frac{16\pi F^2_\pi}{M^2_\pi}\,\cdot\,\frac{1}{1-\frac{(y-2)^2f^2_\pi M^2_\pi}{f^2_dM^2_d}}\nonumber\\
& + \frac{64\pi F^2_\pi}{M^2_\pi}\,\cdot\,\frac{1+\frac{(y-2)f^2_\pi M^2_\pi}{f^2_dM^2_d}+\frac{(y-2)^2f^2_\pi M^4_\pi}{2f^2_dM^4_d}}{\left(1-\frac{(y-2)^2f^2_\pi M^2_\pi}{f^2_d M^2_d}\right)^{2}}.
\label{eq:rpp}
\end{align}
These expressions give the scattering length and effective range in terms of quantities directly measured on the lattice ($F_{\pi}^2$,  $M_{\pi}^2$ and $M_{d}^2$) and two dimensionless fit parameters of the dilaton EFT ($y$ and $f_{\pi}^2/f_d^2$).
We note that the expressions take on simpler forms, derivable from a chiral lagrangian with no dilaton present, when either $f_{\pi}^2/f_d^2 \rightarrow 0$ or $(y-2)\rightarrow 0$.
The first is a limit in which the dilaton decouples from the PNGBs.
The second leaves in place a dilaton contribution to the full scattering amplitude (second line of \eq{eq:fullamp}), but eliminates it for the $I=2$ channel (second and third terms of \eq{eq:i=2dilaton}) at threshold where $s=4M^2_\pi$.


\subsection{Comparison with Lattice Data}

In this section, we first assess qualitatively whether the dilaton EFT is compatible with the lattice data for scattering. We then perform a global fit of all the lattice data to the leading order dilaton EFT, allowing quantitative assessment of the goodness of fit and extraction of the best fit values for the EFT parameters. Finally, we discuss the interpretation of the global fit.

\begin{figure}[t!]
  \centering
  \vspace{12pt}
  \includegraphics[width=0.75\columnwidth]{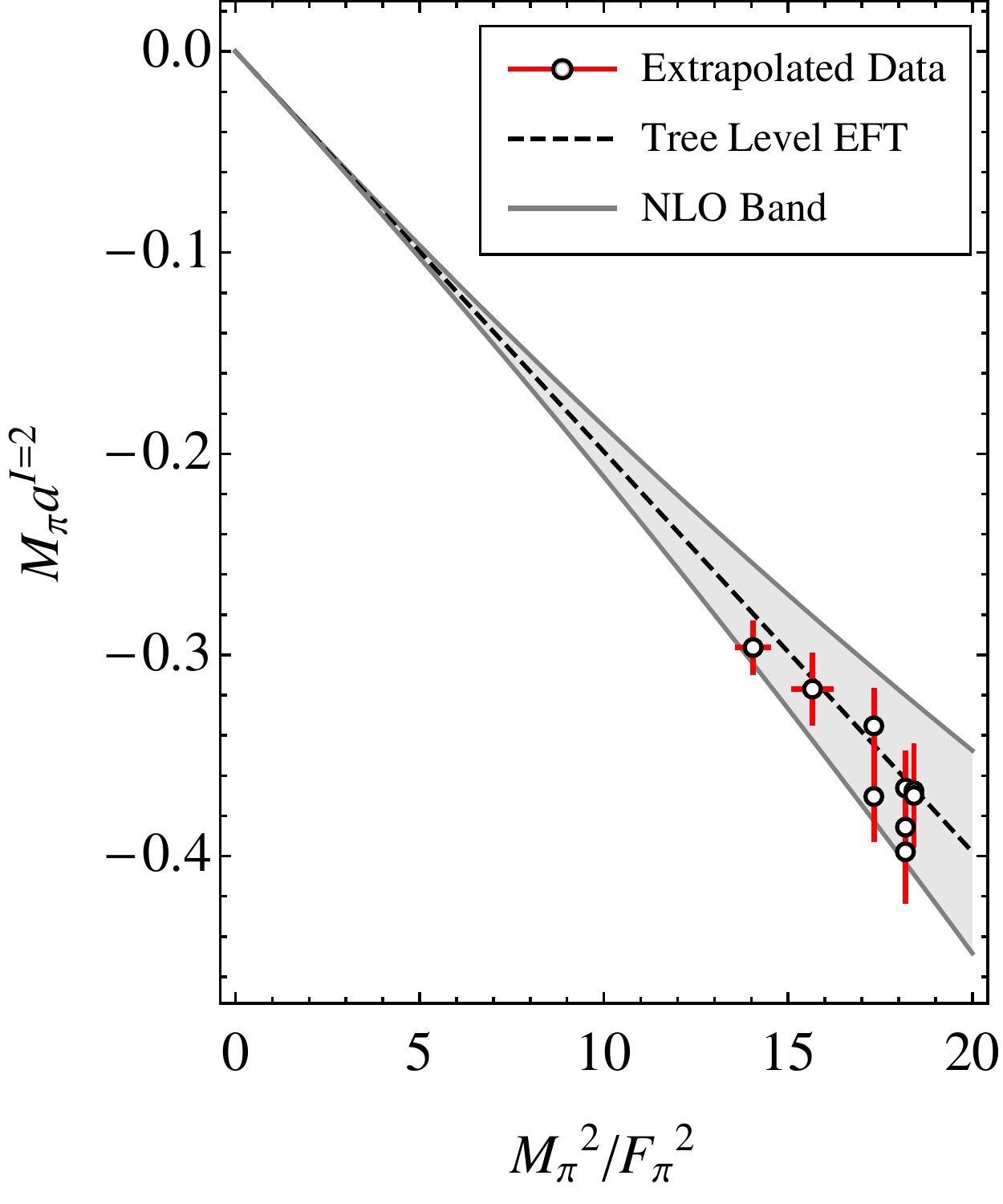}
  \caption{Points representing the scattering lengths determined from lattice data are plotted against the finite--volume extrapolated quantity $M^2_\pi/F^2_\pi$. They are compared with the tree level dilaton EFT prediction given in \protect\eq{eq:dilatonslength}, and shown here as the straight black dashed line. The central values from the global fit are inputted to obtain this dependence. The gray band plotted around the dashed line represents a rough estimate for the size of higher order corrections to the EFT prediction, and is given by Eq.~(\ref{eq:band}).\label{fig:I2datawiththeory}}
\end{figure}

In Fig.~\ref{fig:I2datawiththeory}, we plot the s--wave $I=2$ scattering length against $M^2_\pi/F^2_\pi$. The black dashed line represents the leading order EFT prediction obtained from Eq.~(\ref{eq:dilatonslength}). The gray band represents a rough estimate of the range of values which could be accommodated by higher order corrections to the scattering length in dilaton EFT. The points represent lattice data taken for all the fermion masses and lattice volumes appearing in Table~\ref{tab:data} (9 points total), after extrapolation to the infinite volume and zero scattering momentum limits, needed for direct comparison with the EFT.

To place the points in Fig.~\ref{fig:I2datawiththeory}, we use the infinite volume extrapolated values for $M^2_\pi/F^2_\pi$, shown in Table~\ref{tab:extra}. We also further process the lattice data for $M_\pi/k\cot\delta$ to determine the scattering lengths. This can be viewed as an extrapolation of lattice data taken at nonzero scattering momentum to the zero momentum point. To extract the scattering lengths, we use Eq.~(\ref{eq:effrexp}), along with lattice data for the scattering momenta and the dilaton EFT expression for the effective range shown in Eq.~(\ref{eq:rpp}). Use of the dilaton EFT expressions requires knowledge of parameters, such as $y$ and $f^2_\pi/f^2_d$. We take these values from a global fit of the dilaton EFT to lattice data, which we show in Table~\ref{tab:fitparams}.

The position of the gray band is given by
\begin{align}
a^\text{band}=a^{I=2}\left[1\pm\frac{M^2_\pi}{(4\pi F_\pi)^2}\right],
\label{eq:band}
\end{align}
where $a^{I=2}$ represents the tree level dilaton EFT prediction for the scattering length from Eq.~(\ref{eq:dilatonslength}). The correction term is chosen to scale the same way as a generic next to leading order (NLO) correction with size given by naive dimensional analysis.

In Fig.~\ref{fig:I2datawiththeory}, most but not all of the points overlap with the black dashed line within uncertainties, indicating that the tree level dilaton EFT approximates the data well. However, with the majority of points lying below the dashed line, there is mild evidence of a systematic discrepancy between the tree level theory and lattice data. All of the points lie comfortably within the gray band though, which suggests that small higher order corrections in the dilaton EFT could plausibly improve the agreement.

To quantify the level of agreement between the tree level EFT and the available data, we perform a global fit incorporating the two extrapolated quantities $aM_\pi$ and $aF_\pi$ from Table~\ref{tab:extra}, and the raw lattice data for $aM_d$ and $M_\pi/k\cot\delta$ shown in Table~\ref{tab:data}, where $a$ denotes the lattice spacing. The uncertainties in $aM_\pi$ and $aF_\pi$ are mainly systematic rather than statistical, arising from the use of an approximate model to estimate the size of finite--volume effects, so we treat $aM_\pi$ and $aF_\pi$ as uncorrelated with all other quantities entering the fit. Similarly $aM_d$ measurements are each taken on their own lattice ensembles, and so are uncorrelated with the other fitted quantities. This yields 24 quantities to fit. We also use lattice data for $a^2k^2$, to account for the momentum dependence expected for the scattering phase shift.

We then fit tree-level dilaton EFT expressions to this data set.
First, the EFT determination of $aF_{\pi}$ can be found from solving the transcendental equation in \eq{eq:minimum} and using the scaling relation in \eq{Eq:scaling}.
Then the PNGB and dilaton masses can be determined from Eqs.~(\ref{Eq:Mpiscaling}) and (\ref{eq:Mdscaling}) using this result. The EFT prediction for the scattering phase shift $M_\pi/k\cot\delta$ is made using Eqs.~(\ref{eq:effrexp}), (\ref{eq:dilatonslength}) and (\ref{eq:rpp}) along with data for the scattering momentum.
There are then 6 independent free parameters from the EFT that appear in these expressions. We take them to be $\{y,\,aB_\pi,\,f^2_\pi/f^2_d,\,\Delta,\,a^2f^2_\pi,\,m^2_d/f^2_d\}$.

To do the fit, we construct a chi-square function bilinear in the differences between the four quantities for which we have lattice data and their EFT predictions. Then, we minimize the chi-square function with respect to the 6 EFT parameters. We find that $\chi^2_\text{min}/N_\text{dof}=3.03$ for $N_\text{dof}=24-6=18$, indicating that the tree level EFT is not a perfect description.

The parameters' values at this chi-squared minimum are listed in Table~\ref{tab:fitparams}. We use these values to make the dilaton EFT predictions shown in Figs.~\ref{fig:k2dep}, \ref{fig:I2datawiththeory} and \ref{fig:I0theory}. The values we get lie close to the allowed ranges found in earlier dilaton studies of the $N_f=8$ theory \cite{Appelquist:2019lgk,Golterman:2020tdq} which reported good agreement between the tree level EFT and the available lattice data, but did not consider scattering observables. The estimated uncertainties of the fit parameters are obtained by calculating the inverse Hessian of the chi-squared function at its minimum, to extract the standard errors. These parameter uncertainties are somewhat smaller than those reported in the earlier studies, reflecting the smaller uncertainties in this more recent lattice data set. However these errors, shown for completeness, should be interpreted with caution. As they were extracted from a tree level fit, the best fit values for the EFT parameters may move outside the ranges indicated once higher order effects within the dilaton EFT are accounted for.
\begin{table}[t]
  \centering
  \renewcommand\arraystretch{1.2}  
  \addtolength{\tabcolsep}{4.0 pt} 
  \begin{tabular}{ l l }
    \toprule
    Parameter & Value \\
    \midrule
    $y$ & 2.1321\it{(61)} \\
    $a B_\pi$ & 2.210\it{(60)} \\
    $f^2_\pi/f^2_d$ & 0.0865\it{(42)}\\
    $\Delta$ & 3.11\it{(20)} \\
    $a^2 f^2_\pi$ & $5.8$\it{(2.1)} $\times10^{-5}$ \\
    $m^2_d/f^2_d$ & 1.28\it{(26)} \\
    \bottomrule
  \end{tabular}
  \caption{Best fit values and standard errors for the six dilaton EFT parameters taken from the global fit. These uncertainty ranges do not take into account the effects of higher order contributions arising in the EFT.}
  \label{tab:fitparams}
\end{table}

Even a fundamentally accurate low energy EFT may fail to fit lattice data if the precision of the data exceeds the theoretical error introduced by truncating the EFT at a fixed order in its low energy expansion. This consideration, together with the good qualitative agreement between the tree level EFT and the data suggest that dilaton EFT is a good low energy description of this gauge theory, despite the chi-squared per degree of freedom of the global fit being somewhat larger than one. However, more precise lattice measurements and higher order calculations within the dilaton EFT are now well motivated to confirm this picture.

Our global fit favors $(y-2)^2f^2_\pi/f^2_d\sim 10^{-3}$, ensuring that Eq.~(\ref{eq:dilatonslength}) for the scattering length describes a straight line with gradient $-1/16\pi$ to a high level of approximation. It can be seen in Fig.~\ref{fig:I2datawiththeory} that this relationship is also roughly satisfied by the lattice data. This is also the same prediction for the scattering length as in chiral perturbation theory (without any light scalar state). However we note that the $O(100\%)$ variation of $F_\pi$ with fermion mass evident in the lattice data could not be explained using chiral perturbation theory, and so it would provide a poor global fit to this dataset considered as a whole. The issues encountered when comparing chiral perturbation theory with lattice data for scattering alongside other observables are discussed for  the $N_f=6$ theory in \cite{Appelquist:2012sm} and for the $N_f=8$ theory in \cite{Gasbarro:2017fmi,Gasbarro:2019kgj}.


\subsection{Other Isospin Channels}
In future, scattering lengths in other isospin channels could be measured for the $N_f=8$ theory on the lattice, allowing for new independent tests of this dilaton EFT.
Given that the typical size of scattering momenta obtained on our lattices is much less than the mass splitting between PNGBs in different taste multiplets, 
it makes sense to consider scattering processes involving only the lightest triplet of PNGBs. 
These PNGBs have insufficient kinetic energy to scatter into PNGBs with different tastes for the region of parameter space that we can study on the lattice.

For this lightest triplet of PNGBs, the different possible scattering channels can be fully specified by their $\SU(2)$ isospin quantum numbers.
The dilaton EFT scattering amplitude in the maximal isospin ($I=2$) channels has already been calculated in \eq{eq:i=2dilaton}, and its $\ell=0$ partial wave component shown in \eq{eq:i2l0amp}. The $I=1$ amplitude will have a vanishing $\ell=0$ component due to Bose symmetry, and we shall not consider it further. Finally, the $I=0$ scattering amplitude is a promising target for future lattice study, and we shall calculate the $\ell=0$ scattering length using the dilaton EFT here.

The $I=0$ amplitude has a pole coming from dilaton exchange in the s-channel. It can therefore become divergent for certain physical momentum values, if $M_d>2M_\pi$. However, the EFT still remains weakly coupled in this regime. Also, for scattering near threshold, this s-channel contribution to the amplitude is sensitive to the three-point interaction between a dilaton and PNGB pairs for dilaton momenta $k^2 \sim 4M_{\pi}^2$. This three-point interaction is unsuppressed when $y$ is close to 2, ensuring that the dilaton makes a large contribution to the scattering amplitude near threshold in this channel.

Using \eq{eq:partialwaves} and the $I=0$ equivalents of Eqs.~(\ref{eq:phaseshift}) and (\ref{eq:effrexp}), we can extract the scattering length in the $I=0$ channel
\begin{multline}
  M_\pi a^{I=0} = \frac{7M^2_\pi}{32\pi F^2_\pi}\\+\frac{f^2_\pi}{f^2_d}\frac{M^4_\pi\left(M^2_d(5y^2+4y+20)-8M^2_\pi(y-2)^2\right)}{32\pi F^2_\pi M^2_d\left(M^2_d-4M^2_\pi\right)}.\label{eq:dilatonslength2}
\end{multline}
The first term in this expression is derivable from the chiral lagrangian, and includes no dilaton contribution.
The second term arising from exchanged dilatons vanishes in the limit $f^2_\pi/f^2_d\rightarrow0$ when the dilaton decouples from the PNGBs.
However, unlike in the $I=2$ case, this term remains large when $y-2\ll1$.

\begin{figure}[t!]
  \centering
  \vspace{14pt}
  \includegraphics[width=0.75\columnwidth]{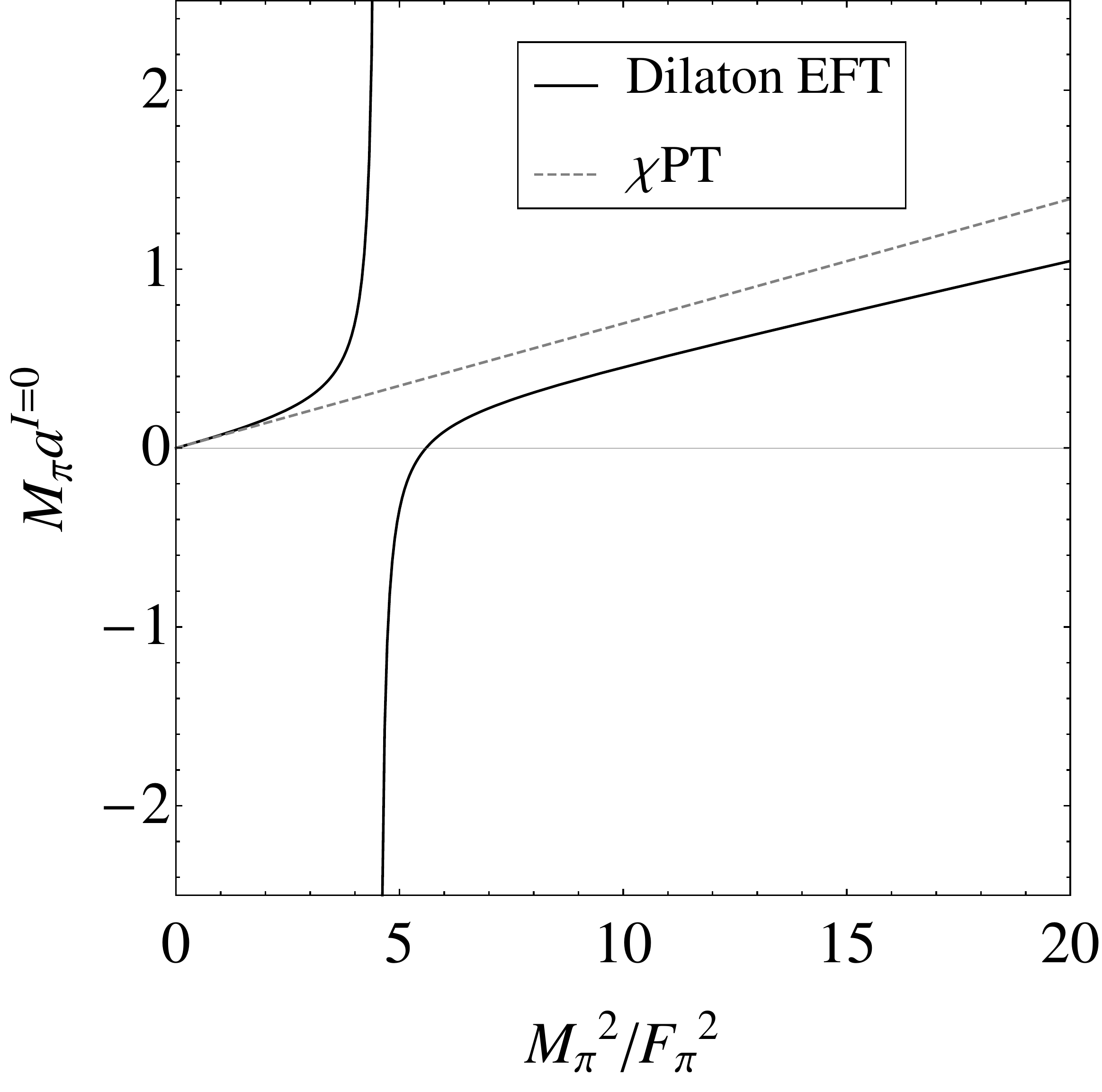}
  \caption{The prediction for the scattering length in the $I=0$ channel, obtained using tree level dilaton EFT from \protect\eq{eq:dilatonslength2}, and the global fit central values is shown here as the black solid line. The gray dashed line representing the prediction from tree level chiral perturbation theory is shown for comparison.\label{fig:I0theory}}
\end{figure}

In \fig{fig:I0theory}, the $I=0$ scattering length predicted from dilaton EFT is plotted.
On the same set of axes, the tree-level chiral lagrangian prediction for the scattering length, corresponding to the first term in \eq{eq:dilatonslength2} is also plotted for reference.
It can be seen that the dilaton EFT prediction for this quantity differs quite significantly from the chiral lagrangian prediction except near the chiral limit, when $M^2_\pi/F^2_\pi$ becomes small, and the PNGBs become significantly lighter than the dilaton.
The spike appearing in the dilaton EFT prediction arises from the s-channel pole in the amplitude.
It occurs when $M^2_d=4M^2_\pi$, which is expected nearer to the chiral limit than current lattice data.

To make the dilaton EFT prediction, central values from the global fit have been used to predict the dependence of $M^2_d/M^2_\pi$ (which appears in \eq{eq:dilatonslength2}) as the chiral limit is approached.
Due to the uncertainties in many of these parameters, there is uncertainty in this prediction away from the regime where lattice data is available $13<M^2_\pi/F^2_\pi<20$.
In particular, the location of the spike depends on the value chosen for $\Delta$.
However, \fig{fig:I0theory} is sufficient to capture the qualitative behavior of the scattering length.

The clear difference between the dilaton EFT and the chiral lagrangian predictions for the $I=0$ scattering length (even at distances from the chiral limit currently accessible on the lattice) make measurement of this quantity a worthwhile goal for future lattice studies. Such a measurement provides a complimentary probe of the interaction strength between the light scalar and the PNGBs, and would further test the dilaton EFT hypothesis.

\section{\label{sec:conclusion}Conclusion}

In this work, we have considered the maximal isospin s-wave scattering of PNGBs in a nearly conformal gauge theory known to possess a light scalar state. We have investigated the possibility that the light scalar, being nearly degenerate with the PNGBs on the gauge ensembles considered, may play a significant role in $\pi\pi$ scattering. While the light scalar pole is not resonant in the $I=2$ channel, the state is exchanged in the T- and U-channels and contributes significantly to the scattering length and effective range if the coupling of the light scalar to the PNGBs in the $I=2$ channel is not too small.

The nonperturbative lattice determination of the scattering phase shift through the L\"uscher procedure is the first such analysis carried out in a theory with a confirmed light composite scalar.
By utilizing moving wall sources and the high gauge statistics available to us, we succeeded in determining the scattering phase shift to high statistical precision.
The calculation was limited by the fact that $(ak)^2$ was small on the available gauge ensembles, leading to a small signal to noise ratio in the difference $E_{\pi\pi} - 2 M_\pi$ compared to $E_{\pi\pi}$ and $M_\pi$ individually.
Nonetheless, our results still showed a significant effective range contribution as demonstrated in Fig.~\ref{fig:k2dep}.
A more detailed study of the effective range contribution would be desirable in future studies by utilizing moving frames and further overlapping volumes.

We compared our data to a leading order dilaton EFT in order to further understand the lattice results and to assess whether the leading order dilaton EFT predictions can provide a good global fit of the data.
We have presented for the first time the leading order expressions for $\pi\pi$ scattering in the dilaton EFT for the $I=2$ and $I=0$ channels.
The complete global fit to the data has revealed slight differences between the predictions of leading order dilaton EFT and the lattice data.

Our result for the scattering length in the $I=0$ channel expressed in \eq{eq:dilatonslength2} and illustrated in Fig.~\ref{fig:I0theory} points to a clear difference between the dilaton EFT and the chiral lagrangian predictions, even at values of $M_\pi^2/F_\pi^2$ currently accessible on the lattice.
While the $I=0$ scattering provides additional numerical challenges on the lattice, these new results from the EFT motivate strongly a dedicated lattice study of this channel.

With these results, we have demonstrated that PNGB scattering provides significant information about the dynamics of nearly conformal gauge theories which exhibit light scalar states.
The scattering phase shifts provide independent tests of effective descriptions such as the dilaton EFT.
In future work, improved precision in the extraction of the scattering phase shifts as well as the exploration of other scattering momenta will provide stronger tests of the dilatonic effective description.
If discrepancies between the tree level dilaton EFT and lattice data persist, this is good motivation for considering NLO terms in the dilatonic effective description.

\begin{acknowledgments}
  R.C.B.~and C.R.~acknowledge United States Department of Energy (DOE) Award No.~{DE-SC0015845}.
  K.C.~acknowledges support from the DOE through the Computational Sciences Graduate Fellowship (DOE CSGF) through grant No.~{DE-SC0019323}.
  G.T.F.~acknowledges support from DOE Award No.~{DE-SC0019061}.
  A.D.G.~is supported by SNSF grant No.~{200021\_17576}.
  A.H.~and E.T.N.~acknowledge support by DOE Award No.~{DE-SC0010005}.
  D.S.~was supported by UK Research and Innovation Future Leader Fellowship No.~{MR/S015418/1} and STFC grant {ST/T000988/1}.
  P.V.~acknowledges the support of the DOE under contract No.~{DE-AC52-07NA27344} (Lawrence Livermore National Laboratory, LLNL).

  We thank the LLNL Multiprogrammatic and Institutional Computing program for Grand Challenge supercomputing allocations. We also thank Argonne Leadership Computing Facility (ALCF) for allocations through the INCITE program. ALCF is supported by DOE contract No.~{DE-AC02-06CH11357}. Computations for this work were carried out in part on facilities of the USQCD Collaboration, which are funded by the Office of Science of the DOE, and on Boston University computers at the MGHPCC, in part funded by the National Science Foundation (award No.~{OCI-1229059}). This research utilized the NVIDIA GPU accelerated Summit supercomputer at Oak Ridge Leadership Computing Facility at the Oak Ridge National Laboratory, which is supported by the Office of Science of the U.S. Department of Energy under Contract No. DE- AC05-00OR22725.
\end{acknowledgments}

\raggedright
\bibliography{I2Scattering}

\begin{thebibliography}{61}%
\makeatletter
\providecommand \@ifxundefined [1]{%
 \@ifx{#1\undefined}
}%
\providecommand \@ifnum [1]{%
 \ifnum #1\expandafter \@firstoftwo
 \else \expandafter \@secondoftwo
 \fi
}%
\providecommand \@ifx [1]{%
 \ifx #1\expandafter \@firstoftwo
 \else \expandafter \@secondoftwo
 \fi
}%
\providecommand \natexlab [1]{#1}%
\providecommand \enquote  [1]{``#1''}%
\providecommand \bibnamefont  [1]{#1}%
\providecommand \bibfnamefont [1]{#1}%
\providecommand \citenamefont [1]{#1}%
\providecommand \href@noop [0]{\@secondoftwo}%
\providecommand \href [0]{\begingroup \@sanitize@url \@href}%
\providecommand \@href[1]{\@@startlink{#1}\@@href}%
\providecommand \@@href[1]{\endgroup#1\@@endlink}%
\providecommand \@sanitize@url [0]{\catcode `\\12\catcode `\$12\catcode
  `\&12\catcode `\#12\catcode `\^12\catcode `\_12\catcode `\%12\relax}%
\providecommand \@@startlink[1]{}%
\providecommand \@@endlink[0]{}%
\providecommand \url  [0]{\begingroup\@sanitize@url \@url }%
\providecommand \@url [1]{\endgroup\@href {#1}{\urlprefix }}%
\providecommand \urlprefix  [0]{URL }%
\providecommand \Eprint [0]{\href }%
\providecommand \doibase [0]{http://dx.doi.org/}%
\providecommand \selectlanguage [0]{\@gobble}%
\providecommand \bibinfo  [0]{\@secondoftwo}%
\providecommand \bibfield  [0]{\@secondoftwo}%
\providecommand \translation [1]{[#1]}%
\providecommand \BibitemOpen [0]{}%
\providecommand \bibitemStop [0]{}%
\providecommand \bibitemNoStop [0]{.\EOS\space}%
\providecommand \EOS [0]{\spacefactor3000\relax}%
\providecommand \BibitemShut  [1]{\csname bibitem#1\endcsname}%
\let\auto@bib@innerbib\@empty
\bibitem [{\citenamefont {Aoki}\ \emph
  {et~al.}(2014{\natexlab{a}})\citenamefont {Aoki}, \citenamefont {Aoyama},
  \citenamefont {Kurachi}, \citenamefont {Maskawa}, \citenamefont {Nagai},
  \citenamefont {Ohki}, \citenamefont {Rinaldi}, \citenamefont {Shibata},
  \citenamefont {Yamawaki},\ and\ \citenamefont {Yamazaki}}]{Aoki:2013pca}%
  \BibitemOpen
  \bibfield  {author} {\bibinfo {author} {\bibfnamefont {Y.}~\bibnamefont
  {Aoki}}, \bibinfo {author} {\bibfnamefont {T.}~\bibnamefont {Aoyama}},
  \bibinfo {author} {\bibfnamefont {M.}~\bibnamefont {Kurachi}}, \bibinfo
  {author} {\bibfnamefont {T.}~\bibnamefont {Maskawa}}, \bibinfo {author}
  {\bibfnamefont {K.-i.}\ \bibnamefont {Nagai}}, \bibinfo {author}
  {\bibfnamefont {H.}~\bibnamefont {Ohki}}, \bibinfo {author} {\bibfnamefont
  {Enrico}\ \bibnamefont {Rinaldi}}, \bibinfo {author} {\bibfnamefont
  {A.}~\bibnamefont {Shibata}}, \bibinfo {author} {\bibfnamefont
  {K.}~\bibnamefont {Yamawaki}}, \ and\ \bibinfo {author} {\bibfnamefont
  {T.}~\bibnamefont {Yamazaki}} (\bibinfo {collaboration} {LatKMI
  Collaboration}),\ }\bibfield  {title} {\enquote {\bibinfo {title} {{The
  scalar spectrum of many-flavour QCD}},}\ }in\ \href {\doibase
  10.1142/9789814566254_0009} {\emph {\bibinfo {booktitle} {{Strong Coupling
  Gauge Theories in the LHC Perspective (SCGT12)}}}}\ (\bibinfo {year} {2014})\
  pp.\ \bibinfo {pages} {80--88},\ \Eprint {http://arxiv.org/abs/1302.4577}
  {arXiv:1302.4577} \BibitemShut {NoStop}%
\bibitem [{\citenamefont {Aoki}\ \emph {et~al.}(2013)\citenamefont {Aoki},
  \citenamefont {Aoyama}, \citenamefont {Kurachi}, \citenamefont {Maskawa},
  \citenamefont {Nagai}, \citenamefont {Ohki}, \citenamefont {Rinaldi},
  \citenamefont {Shibata}, \citenamefont {Yamawaki},\ and\ \citenamefont
  {Yamazaki}}]{Aoki:2013zsa}%
  \BibitemOpen
  \bibfield  {author} {\bibinfo {author} {\bibfnamefont {Y.}~\bibnamefont
  {Aoki}}, \bibinfo {author} {\bibfnamefont {T.}~\bibnamefont {Aoyama}},
  \bibinfo {author} {\bibfnamefont {M.}~\bibnamefont {Kurachi}}, \bibinfo
  {author} {\bibfnamefont {T.}~\bibnamefont {Maskawa}}, \bibinfo {author}
  {\bibfnamefont {K.-i.}\ \bibnamefont {Nagai}}, \bibinfo {author}
  {\bibfnamefont {H.}~\bibnamefont {Ohki}}, \bibinfo {author} {\bibfnamefont
  {E.}~\bibnamefont {Rinaldi}}, \bibinfo {author} {\bibfnamefont
  {A.}~\bibnamefont {Shibata}}, \bibinfo {author} {\bibfnamefont
  {K.}~\bibnamefont {Yamawaki}}, \ and\ \bibinfo {author} {\bibfnamefont
  {T.}~\bibnamefont {Yamazaki}} (\bibinfo {collaboration} {LatKMI
  Collaboration}),\ }\bibfield  {title} {\enquote {\bibinfo {title} {{Light
  composite scalar in twelve-flavor QCD on the lattice}},}\ }\href {\doibase
  10.1103/PhysRevLett.111.162001} {\bibfield  {journal} {\bibinfo  {journal}
  {Phys. Rev. Lett.}\ }\textbf {\bibinfo {volume} {111}},\ \bibinfo {pages}
  {162001} (\bibinfo {year} {2013})},\ \Eprint {http://arxiv.org/abs/1305.6006}
  {arXiv:1305.6006} \BibitemShut {NoStop}%
\bibitem [{\citenamefont {Aoki}\ \emph
  {et~al.}(2014{\natexlab{b}})\citenamefont {Aoki}, \citenamefont {Aoyama},
  \citenamefont {Kurachi}, \citenamefont {Maskawa}, \citenamefont {Miura},
  \citenamefont {Nagai}, \citenamefont {Ohki}, \citenamefont {Rinaldi},
  \citenamefont {Shibata}, \citenamefont {Yamawaki},\ and\ \citenamefont
  {Yamazaki}}]{Aoki:2013hla}%
  \BibitemOpen
  \bibfield  {author} {\bibinfo {author} {\bibfnamefont {Y.}~\bibnamefont
  {Aoki}}, \bibinfo {author} {\bibfnamefont {T.}~\bibnamefont {Aoyama}},
  \bibinfo {author} {\bibfnamefont {M.}~\bibnamefont {Kurachi}}, \bibinfo
  {author} {\bibfnamefont {T.}~\bibnamefont {Maskawa}}, \bibinfo {author}
  {\bibfnamefont {K.}~\bibnamefont {Miura}}, \bibinfo {author} {\bibfnamefont
  {K.-i.}\ \bibnamefont {Nagai}}, \bibinfo {author} {\bibfnamefont
  {H.}~\bibnamefont {Ohki}}, \bibinfo {author} {\bibfnamefont {E.}~\bibnamefont
  {Rinaldi}}, \bibinfo {author} {\bibfnamefont {A.}~\bibnamefont {Shibata}},
  \bibinfo {author} {\bibfnamefont {K.}~\bibnamefont {Yamawaki}}, \ and\
  \bibinfo {author} {\bibfnamefont {T.}~\bibnamefont {Yamazaki}} (\bibinfo
  {collaboration} {LatKMI Collaboration}),\ }\bibfield  {title} {\enquote
  {\bibinfo {title} {{Composite flavor-singlet scalar in twelve-flavor QCD}},}\
  }\href {\doibase 10.22323/1.187.0077} {\bibfield  {journal} {\bibinfo
  {journal} {Proc. Sci.}\ }\textbf {\bibinfo {volume} {LATTICE2013}},\ \bibinfo
  {pages} {077} (\bibinfo {year} {2014}{\natexlab{b}})},\ \Eprint
  {http://arxiv.org/abs/1311.6885} {arXiv:1311.6885} \BibitemShut {NoStop}%
\bibitem [{\citenamefont {Aoki}\ \emph
  {et~al.}(2014{\natexlab{c}})\citenamefont {Aoki}, \citenamefont {Aoyama},
  \citenamefont {Kurachi}, \citenamefont {Maskawa}, \citenamefont {Miura},
  \citenamefont {Nagai}, \citenamefont {Ohki}, \citenamefont {Rinaldi},
  \citenamefont {Shibata}, \citenamefont {Yamawaki},\ and\ \citenamefont
  {Yamazaki}}]{Aoki:2014oha}%
  \BibitemOpen
  \bibfield  {author} {\bibinfo {author} {\bibfnamefont {Y.}~\bibnamefont
  {Aoki}}, \bibinfo {author} {\bibfnamefont {T.}~\bibnamefont {Aoyama}},
  \bibinfo {author} {\bibfnamefont {M.}~\bibnamefont {Kurachi}}, \bibinfo
  {author} {\bibfnamefont {T.}~\bibnamefont {Maskawa}}, \bibinfo {author}
  {\bibfnamefont {K.}~\bibnamefont {Miura}}, \bibinfo {author} {\bibfnamefont
  {K.-i.}\ \bibnamefont {Nagai}}, \bibinfo {author} {\bibfnamefont
  {H.}~\bibnamefont {Ohki}}, \bibinfo {author} {\bibfnamefont {E.}~\bibnamefont
  {Rinaldi}}, \bibinfo {author} {\bibfnamefont {A.}~\bibnamefont {Shibata}},
  \bibinfo {author} {\bibfnamefont {K.}~\bibnamefont {Yamawaki}}, \ and\
  \bibinfo {author} {\bibfnamefont {T.}~\bibnamefont {Yamazaki}} (\bibinfo
  {collaboration} {LatKMI Collaboration}),\ }\bibfield  {title} {\enquote
  {\bibinfo {title} {{Light composite scalar in eight-flavor QCD on the
  lattice}},}\ }\href {\doibase 10.1103/PhysRevD.89.111502} {\bibfield
  {journal} {\bibinfo  {journal} {Phys. Rev. D}\ }\textbf {\bibinfo {volume}
  {89}},\ \bibinfo {pages} {111502} (\bibinfo {year} {2014}{\natexlab{c}})},\
  \Eprint {http://arxiv.org/abs/1403.5000} {arXiv:1403.5000} \BibitemShut
  {NoStop}%
\bibitem [{\citenamefont {Aoki}\ \emph {et~al.}(2017)\citenamefont {Aoki},
  \citenamefont {Aoyama}, \citenamefont {Bennett}, \citenamefont {Kurachi},
  \citenamefont {Maskawa}, \citenamefont {Miura}, \citenamefont {Nagai},
  \citenamefont {Ohki}, \citenamefont {Rinaldi}, \citenamefont {Shibata},
  \citenamefont {Yamawaki},\ and\ \citenamefont {Yamazaki}}]{Aoki:2016wnc}%
  \BibitemOpen
  \bibfield  {author} {\bibinfo {author} {\bibfnamefont {Y.}~\bibnamefont
  {Aoki}}, \bibinfo {author} {\bibfnamefont {T.}~\bibnamefont {Aoyama}},
  \bibinfo {author} {\bibfnamefont {Ed}~\bibnamefont {Bennett}}, \bibinfo
  {author} {\bibfnamefont {M.}~\bibnamefont {Kurachi}}, \bibinfo {author}
  {\bibfnamefont {T.}~\bibnamefont {Maskawa}}, \bibinfo {author} {\bibfnamefont
  {K.}~\bibnamefont {Miura}}, \bibinfo {author} {\bibfnamefont {K.-i.}\
  \bibnamefont {Nagai}}, \bibinfo {author} {\bibfnamefont {H.}~\bibnamefont
  {Ohki}}, \bibinfo {author} {\bibfnamefont {E.}~\bibnamefont {Rinaldi}},
  \bibinfo {author} {\bibfnamefont {A.}~\bibnamefont {Shibata}}, \bibinfo
  {author} {\bibfnamefont {K.}~\bibnamefont {Yamawaki}}, \ and\ \bibinfo
  {author} {\bibfnamefont {T.}~\bibnamefont {Yamazaki}} (\bibinfo
  {collaboration} {LatKMI Collaboration}),\ }\bibfield  {title} {\enquote
  {\bibinfo {title} {{Light flavor-singlet scalars and walking signals in
  $N_f=8$ QCD on the lattice}},}\ }\href {\doibase 10.1103/PhysRevD.96.014508}
  {\bibfield  {journal} {\bibinfo  {journal} {Phys. Rev. D}\ }\textbf {\bibinfo
  {volume} {96}},\ \bibinfo {pages} {014508} (\bibinfo {year} {2017})},\
  \Eprint {http://arxiv.org/abs/1610.07011} {arXiv:1610.07011} \BibitemShut
  {NoStop}%
\bibitem [{\citenamefont {Athenodorou}\ \emph {et~al.}(2015)\citenamefont
  {Athenodorou}, \citenamefont {Bennett}, \citenamefont {Bergner},\ and\
  \citenamefont {Lucini}}]{Athenodorou:2014eua}%
  \BibitemOpen
  \bibfield  {author} {\bibinfo {author} {\bibfnamefont {A.}~\bibnamefont
  {Athenodorou}}, \bibinfo {author} {\bibfnamefont {E.}~\bibnamefont
  {Bennett}}, \bibinfo {author} {\bibfnamefont {G.}~\bibnamefont {Bergner}}, \
  and\ \bibinfo {author} {\bibfnamefont {B.}~\bibnamefont {Lucini}},\
  }\bibfield  {title} {\enquote {\bibinfo {title} {{Infrared regime of SU(2)
  with one adjoint Dirac flavor}},}\ }\href {\doibase
  10.1103/PhysRevD.91.114508} {\bibfield  {journal} {\bibinfo  {journal} {Phys.
  Rev. D}\ }\textbf {\bibinfo {volume} {91}},\ \bibinfo {pages} {114508}
  (\bibinfo {year} {2015})},\ \Eprint {http://arxiv.org/abs/1412.5994}
  {arXiv:1412.5994} \BibitemShut {NoStop}%
\bibitem [{\citenamefont {Fodor}\ \emph {et~al.}(2015)\citenamefont {Fodor},
  \citenamefont {Holland}, \citenamefont {Kuti}, \citenamefont {Mondal},
  \citenamefont {Nogradi},\ and\ \citenamefont {Wong}}]{Fodor:2015vwa}%
  \BibitemOpen
  \bibfield  {author} {\bibinfo {author} {\bibfnamefont {Z.}~\bibnamefont
  {Fodor}}, \bibinfo {author} {\bibfnamefont {K.}~\bibnamefont {Holland}},
  \bibinfo {author} {\bibfnamefont {J.}~\bibnamefont {Kuti}}, \bibinfo {author}
  {\bibfnamefont {S.}~\bibnamefont {Mondal}}, \bibinfo {author} {\bibfnamefont
  {D.}~\bibnamefont {Nogradi}}, \ and\ \bibinfo {author} {\bibfnamefont
  {C.~H.}\ \bibnamefont {Wong}},\ }\bibfield  {title} {\enquote {\bibinfo
  {title} {{Toward the minimal realization of a light composite Higgs}},}\
  }\href {\doibase 10.22323/1.214.0244} {\bibfield  {journal} {\bibinfo
  {journal} {Proc. Sci.}\ }\textbf {\bibinfo {volume} {LATTICE2014}},\ \bibinfo
  {pages} {244} (\bibinfo {year} {2015})},\ \Eprint
  {http://arxiv.org/abs/1502.00028} {arXiv:1502.00028} \BibitemShut {NoStop}%
\bibitem [{\citenamefont {Rinaldi}(2017)}]{Rinaldi:2015axa}%
  \BibitemOpen
  \bibfield  {author} {\bibinfo {author} {\bibfnamefont {E.}~\bibnamefont
  {Rinaldi}} (\bibinfo {collaboration} {LSD Collaboration}),\ }\bibfield
  {title} {\enquote {\bibinfo {title} {{Investigation of the scalar spectrum in
  SU(3) with eight degenerate flavors}},}\ }\href {\doibase
  10.1142/S0217751X17470029} {\bibfield  {journal} {\bibinfo  {journal} {Int.
  J. Mod. Phys. A}\ }\textbf {\bibinfo {volume} {32}},\ \bibinfo {pages}
  {1747002} (\bibinfo {year} {2017})},\ \Eprint
  {http://arxiv.org/abs/1510.06771} {arXiv:1510.06771} \BibitemShut {NoStop}%
\bibitem [{\citenamefont {Brower}\ \emph {et~al.}(2016)\citenamefont {Brower},
  \citenamefont {Hasenfratz}, \citenamefont {Rebbi}, \citenamefont {Weinberg},\
  and\ \citenamefont {Witzel}}]{Brower:2015owo}%
  \BibitemOpen
  \bibfield  {author} {\bibinfo {author} {\bibfnamefont {R.~C.}\ \bibnamefont
  {Brower}}, \bibinfo {author} {\bibfnamefont {A.}~\bibnamefont {Hasenfratz}},
  \bibinfo {author} {\bibfnamefont {C.}~\bibnamefont {Rebbi}}, \bibinfo
  {author} {\bibfnamefont {E.}~\bibnamefont {Weinberg}}, \ and\ \bibinfo
  {author} {\bibfnamefont {O.}~\bibnamefont {Witzel}},\ }\bibfield  {title}
  {\enquote {\bibinfo {title} {{Composite Higgs model at a conformal fixed
  point}},}\ }\href {\doibase 10.1103/PhysRevD.93.075028} {\bibfield  {journal}
  {\bibinfo  {journal} {Phys. Rev. D}\ }\textbf {\bibinfo {volume} {93}},\
  \bibinfo {pages} {075028} (\bibinfo {year} {2016})},\ \Eprint
  {http://arxiv.org/abs/1512.02576} {arXiv:1512.02576} \BibitemShut {NoStop}%
\bibitem [{\citenamefont {Fodor}\ \emph {et~al.}(2016)\citenamefont {Fodor},
  \citenamefont {Holland}, \citenamefont {Kuti}, \citenamefont {Mondal},
  \citenamefont {Nogradi},\ and\ \citenamefont {Wong}}]{Fodor:2016pls}%
  \BibitemOpen
  \bibfield  {author} {\bibinfo {author} {\bibfnamefont {Z.}~\bibnamefont
  {Fodor}}, \bibinfo {author} {\bibfnamefont {K.}~\bibnamefont {Holland}},
  \bibinfo {author} {\bibfnamefont {J.}~\bibnamefont {Kuti}}, \bibinfo {author}
  {\bibfnamefont {S.}~\bibnamefont {Mondal}}, \bibinfo {author} {\bibfnamefont
  {D.}~\bibnamefont {Nogradi}}, \ and\ \bibinfo {author} {\bibfnamefont
  {C.~H.}\ \bibnamefont {Wong}},\ }\bibfield  {title} {\enquote {\bibinfo
  {title} {{Status of a minimal composite Higgs theory}},}\ }\href {\doibase
  10.22323/1.251.0219} {\bibfield  {journal} {\bibinfo  {journal} {Proc. Sci.}\
  }\textbf {\bibinfo {volume} {LATTICE2015}},\ \bibinfo {pages} {219} (\bibinfo
  {year} {2016})},\ \Eprint {http://arxiv.org/abs/1605.08750}
  {arXiv:1605.08750} \BibitemShut {NoStop}%
\bibitem [{\citenamefont {Hasenfratz}\ \emph {et~al.}(2017)\citenamefont
  {Hasenfratz}, \citenamefont {Rebbi},\ and\ \citenamefont
  {Witzel}}]{Hasenfratz:2016gut}%
  \BibitemOpen
  \bibfield  {author} {\bibinfo {author} {\bibfnamefont {A.}~\bibnamefont
  {Hasenfratz}}, \bibinfo {author} {\bibfnamefont {C.}~\bibnamefont {Rebbi}}, \
  and\ \bibinfo {author} {\bibfnamefont {O.}~\bibnamefont {Witzel}},\
  }\bibfield  {title} {\enquote {\bibinfo {title} {{Large scale separation and
  resonances within LHC range from a prototype BSM model}},}\ }\href {\doibase
  10.1016/j.physletb.2017.07.058} {\bibfield  {journal} {\bibinfo  {journal}
  {Phys. Lett. B}\ }\textbf {\bibinfo {volume} {773}},\ \bibinfo {pages}
  {86--90} (\bibinfo {year} {2017})},\ \Eprint
  {http://arxiv.org/abs/1609.01401} {arXiv:1609.01401} \BibitemShut {NoStop}%
\bibitem [{\citenamefont {Del~Debbio}\ \emph {et~al.}(2016)\citenamefont
  {Del~Debbio}, \citenamefont {Lucini}, \citenamefont {Patella}, \citenamefont
  {Pica},\ and\ \citenamefont {Rago}}]{DelDebbio:2015byq}%
  \BibitemOpen
  \bibfield  {author} {\bibinfo {author} {\bibfnamefont {L.}~\bibnamefont
  {Del~Debbio}}, \bibinfo {author} {\bibfnamefont {B.}~\bibnamefont {Lucini}},
  \bibinfo {author} {\bibfnamefont {A.}~\bibnamefont {Patella}}, \bibinfo
  {author} {\bibfnamefont {C.}~\bibnamefont {Pica}}, \ and\ \bibinfo {author}
  {\bibfnamefont {A.}~\bibnamefont {Rago}},\ }\bibfield  {title} {\enquote
  {\bibinfo {title} {{Large volumes and spectroscopy of walking theories}},}\
  }\href {\doibase 10.1103/PhysRevD.93.054505} {\bibfield  {journal} {\bibinfo
  {journal} {Phys. Rev. D}\ }\textbf {\bibinfo {volume} {93}},\ \bibinfo
  {pages} {054505} (\bibinfo {year} {2016})},\ \Eprint
  {http://arxiv.org/abs/1512.08242} {arXiv:1512.08242} \BibitemShut {NoStop}%
\bibitem [{\citenamefont {Appelquist}\ \emph {et~al.}(2016)\citenamefont
  {Appelquist}, \citenamefont {Brower}, \citenamefont {Fleming}, \citenamefont
  {Hasenfratz}, \citenamefont {Jin}, \citenamefont {Kiskis}, \citenamefont
  {Osborn}, \citenamefont {Rebbi}, \citenamefont {Rinaldi}, \citenamefont
  {Schaich}, \citenamefont {Vranas}, \citenamefont {Weinberg},\ and\
  \citenamefont {Witzel}}]{Appelquist:2016viq}%
  \BibitemOpen
  \bibfield  {author} {\bibinfo {author} {\bibfnamefont {T.}~\bibnamefont
  {Appelquist}}, \bibinfo {author} {\bibfnamefont {R.~C.}\ \bibnamefont
  {Brower}}, \bibinfo {author} {\bibfnamefont {G.~T.}\ \bibnamefont {Fleming}},
  \bibinfo {author} {\bibfnamefont {A.}~\bibnamefont {Hasenfratz}}, \bibinfo
  {author} {\bibfnamefont {X.-Y.}\ \bibnamefont {Jin}}, \bibinfo {author}
  {\bibfnamefont {E.~T.}\ \bibnamefont {Kiskis}, \bibfnamefont {J.~Neil}},
  \bibinfo {author} {\bibfnamefont {J.~C.}\ \bibnamefont {Osborn}}, \bibinfo
  {author} {\bibfnamefont {C.}~\bibnamefont {Rebbi}}, \bibinfo {author}
  {\bibfnamefont {E.}~\bibnamefont {Rinaldi}}, \bibinfo {author} {\bibfnamefont
  {D.}~\bibnamefont {Schaich}}, \bibinfo {author} {\bibfnamefont
  {P.}~\bibnamefont {Vranas}}, \bibinfo {author} {\bibfnamefont
  {E.}~\bibnamefont {Weinberg}}, \ and\ \bibinfo {author} {\bibfnamefont
  {O.}~\bibnamefont {Witzel}} (\bibinfo {collaboration} {LSD Collaboration}),\
  }\bibfield  {title} {\enquote {\bibinfo {title} {{Strongly interacting
  dynamics and the search for new physics at the LHC}},}\ }\href {\doibase
  10.1103/PhysRevD.93.114514} {\bibfield  {journal} {\bibinfo  {journal} {Phys.
  Rev. D}\ }\textbf {\bibinfo {volume} {93}},\ \bibinfo {pages} {114514}
  (\bibinfo {year} {2016})},\ \Eprint {http://arxiv.org/abs/1601.04027}
  {arXiv:1601.04027} \BibitemShut {NoStop}%
\bibitem [{\citenamefont {Appelquist}\ \emph {et~al.}(2019)\citenamefont
  {Appelquist}, \citenamefont {Brower}, \citenamefont {Fleming}, \citenamefont
  {Gasbarro}, \citenamefont {Hasenfratz}, \citenamefont {Jin}, \citenamefont
  {Neil}, \citenamefont {Osborn}, \citenamefont {Rebbi}, \citenamefont
  {Rinaldi}, \citenamefont {Schaich}, \citenamefont {Vranas}, \citenamefont
  {Weinberg},\ and\ \citenamefont {Witzel}}]{Appelquist:2018yqe}%
  \BibitemOpen
  \bibfield  {author} {\bibinfo {author} {\bibfnamefont {T.}~\bibnamefont
  {Appelquist}}, \bibinfo {author} {\bibfnamefont {R.~C.}\ \bibnamefont
  {Brower}}, \bibinfo {author} {\bibfnamefont {G.~T.}\ \bibnamefont {Fleming}},
  \bibinfo {author} {\bibfnamefont {A.}~\bibnamefont {Gasbarro}}, \bibinfo
  {author} {\bibfnamefont {A.}~\bibnamefont {Hasenfratz}}, \bibinfo {author}
  {\bibfnamefont {X.-Y.}\ \bibnamefont {Jin}}, \bibinfo {author} {\bibfnamefont
  {E.~T.}\ \bibnamefont {Neil}}, \bibinfo {author} {\bibfnamefont {J.~C.}\
  \bibnamefont {Osborn}}, \bibinfo {author} {\bibfnamefont {C.}~\bibnamefont
  {Rebbi}}, \bibinfo {author} {\bibfnamefont {E.}~\bibnamefont {Rinaldi}},
  \bibinfo {author} {\bibfnamefont {D.}~\bibnamefont {Schaich}}, \bibinfo
  {author} {\bibfnamefont {P.}~\bibnamefont {Vranas}}, \bibinfo {author}
  {\bibfnamefont {E.}~\bibnamefont {Weinberg}}, \ and\ \bibinfo {author}
  {\bibfnamefont {O.}~\bibnamefont {Witzel}} (\bibinfo {collaboration} {LSD
  Collaboration}),\ }\bibfield  {title} {\enquote {\bibinfo {title}
  {{Nonperturbative investigations of SU(3) gauge theory with eight dynamical
  flavors}},}\ }\href {\doibase 10.1103/PhysRevD.99.014509} {\bibfield
  {journal} {\bibinfo  {journal} {Phys. Rev. D}\ }\textbf {\bibinfo {volume}
  {99}},\ \bibinfo {pages} {014509} (\bibinfo {year} {2019})},\ \Eprint
  {http://arxiv.org/abs/1807.08411} {arXiv:1807.08411} \BibitemShut {NoStop}%
\bibitem [{\citenamefont {Gasbarro}\ and\ \citenamefont
  {Fleming}(2017)}]{Gasbarro:2017fmi}%
  \BibitemOpen
  \bibfield  {author} {\bibinfo {author} {\bibfnamefont {A.~D.}\ \bibnamefont
  {Gasbarro}}\ and\ \bibinfo {author} {\bibfnamefont {G.~T.}\ \bibnamefont
  {Fleming}},\ }\bibfield  {title} {\enquote {\bibinfo {title} {{Examining the
  Low Energy Dynamics of Walking Gauge Theory}},}\ }\href {\doibase
  10.22323/1.256.0242} {\bibfield  {journal} {\bibinfo  {journal} {Proc. Sci.}\
  }\textbf {\bibinfo {volume} {LATTICE2016}},\ \bibinfo {pages} {242} (\bibinfo
  {year} {2017})},\ \Eprint {http://arxiv.org/abs/1702.00480}
  {arXiv:1702.00480} \BibitemShut {NoStop}%
\bibitem [{\citenamefont {Gasbarro}(2019)}]{Gasbarro:2019kgj}%
  \BibitemOpen
  \bibfield  {author} {\bibinfo {author} {\bibfnamefont {Andrew~David}\
  \bibnamefont {Gasbarro}},\ }\emph {\bibinfo {title} {{Studies of Conformal
  Behavior in Strongly Interacting Quantum Field Theories}}},\ \href@noop {}
  {\bibinfo {type} {Doctoral thesis}},\ \bibinfo  {school} {Yale} (\bibinfo
  {year} {2019}),\ \Eprint {http://arxiv.org/abs/1911.00442} {arXiv:1911.00442
  [hep-lat]} \BibitemShut {NoStop}%
\bibitem [{\citenamefont {Athenodorou}\ \emph {et~al.}(2017)\citenamefont
  {Athenodorou}, \citenamefont {Bennett}, \citenamefont {Bergner},
  \citenamefont {Elander}, \citenamefont {Lin}, \citenamefont {Lucini},\ and\
  \citenamefont {Piai}}]{Athenodorou:2017dbf}%
  \BibitemOpen
  \bibfield  {author} {\bibinfo {author} {\bibfnamefont {A.}~\bibnamefont
  {Athenodorou}}, \bibinfo {author} {\bibfnamefont {E.}~\bibnamefont
  {Bennett}}, \bibinfo {author} {\bibfnamefont {G.}~\bibnamefont {Bergner}},
  \bibinfo {author} {\bibfnamefont {D.}~\bibnamefont {Elander}}, \bibinfo
  {author} {\bibfnamefont {C.-J.~D.}\ \bibnamefont {Lin}}, \bibinfo {author}
  {\bibfnamefont {B.}~\bibnamefont {Lucini}}, \ and\ \bibinfo {author}
  {\bibfnamefont {M.}~\bibnamefont {Piai}},\ }\bibfield  {title} {\enquote
  {\bibinfo {title} {{Large mass hierarchies from strongly-coupled
  dynamics}},}\ }\href {\doibase 10.22323/1.256.0232} {\bibfield  {journal}
  {\bibinfo  {journal} {Proc. Sci.}\ }\textbf {\bibinfo {volume}
  {LATTICE2016}},\ \bibinfo {pages} {232} (\bibinfo {year} {2017})},\ \Eprint
  {http://arxiv.org/abs/1702.06452} {arXiv:1702.06452} \BibitemShut {NoStop}%
\bibitem [{\citenamefont {Fodor}\ \emph {et~al.}(2018)\citenamefont {Fodor},
  \citenamefont {Holland}, \citenamefont {Kuti}, \citenamefont {Nogradi},\ and\
  \citenamefont {Wong}}]{Fodor:2017nlp}%
  \BibitemOpen
  \bibfield  {author} {\bibinfo {author} {\bibfnamefont {Z.}~\bibnamefont
  {Fodor}}, \bibinfo {author} {\bibfnamefont {K.}~\bibnamefont {Holland}},
  \bibinfo {author} {\bibfnamefont {J.}~\bibnamefont {Kuti}}, \bibinfo {author}
  {\bibfnamefont {D.}~\bibnamefont {Nogradi}}, \ and\ \bibinfo {author}
  {\bibfnamefont {C.~H.}\ \bibnamefont {Wong}},\ }\bibfield  {title} {\enquote
  {\bibinfo {title} {{The twelve-flavor $\beta$-function and dilaton tests of
  the sextet scalar}},}\ }\href {\doibase 10.1051/epjconf/201817508015}
  {\bibfield  {journal} {\bibinfo  {journal} {EPJ Web Conf.}\ }\textbf
  {\bibinfo {volume} {175}},\ \bibinfo {pages} {08015} (\bibinfo {year}
  {2018})},\ \Eprint {http://arxiv.org/abs/1712.08594} {arXiv:1712.08594}
  \BibitemShut {NoStop}%
\bibitem [{\citenamefont {Athenodorou}\ \emph {et~al.}(2021)\citenamefont
  {Athenodorou}, \citenamefont {Bennett}, \citenamefont {Bergner},\ and\
  \citenamefont {Lucini}}]{Athenodorou:2021wom}%
  \BibitemOpen
  \bibfield  {author} {\bibinfo {author} {\bibfnamefont {A.}~\bibnamefont
  {Athenodorou}}, \bibinfo {author} {\bibfnamefont {E.}~\bibnamefont
  {Bennett}}, \bibinfo {author} {\bibfnamefont {G.}~\bibnamefont {Bergner}}, \
  and\ \bibinfo {author} {\bibfnamefont {B.}~\bibnamefont {Lucini}},\
  }\bibfield  {title} {\enquote {\bibinfo {title} {{Investigating the conformal
  behaviour of SU(2) with one adjoint Dirac flavor}},}\ }\href@noop {} {\
  (\bibinfo {year} {2021})},\ \Eprint {http://arxiv.org/abs/2103.10485}
  {arXiv:2103.10485} \BibitemShut {NoStop}%
\bibitem [{\citenamefont {Brower}\ \emph {et~al.}(2019)\citenamefont {Brower},
  \citenamefont {Hasenfratz}, \citenamefont {Neil}, \citenamefont {Catterall},
  \citenamefont {Fleming}, \citenamefont {Giedt}, \citenamefont {Rinaldi},
  \citenamefont {Schaich}, \citenamefont {Weinberg},\ and\ \citenamefont
  {Witzel}}]{Brower:2019oor}%
  \BibitemOpen
  \bibfield  {author} {\bibinfo {author} {\bibfnamefont {R.~C.}\ \bibnamefont
  {Brower}}, \bibinfo {author} {\bibfnamefont {A.}~\bibnamefont {Hasenfratz}},
  \bibinfo {author} {\bibfnamefont {E.~T.}\ \bibnamefont {Neil}}, \bibinfo
  {author} {\bibfnamefont {S.}~\bibnamefont {Catterall}}, \bibinfo {author}
  {\bibfnamefont {G.}~\bibnamefont {Fleming}}, \bibinfo {author} {\bibfnamefont
  {J.}~\bibnamefont {Giedt}}, \bibinfo {author} {\bibfnamefont
  {E.}~\bibnamefont {Rinaldi}}, \bibinfo {author} {\bibfnamefont
  {D.}~\bibnamefont {Schaich}}, \bibinfo {author} {\bibfnamefont
  {E.}~\bibnamefont {Weinberg}}, \ and\ \bibinfo {author} {\bibfnamefont
  {O.}~\bibnamefont {Witzel}} (\bibinfo {collaboration} {USQCD}),\ }\bibfield
  {title} {\enquote {\bibinfo {title} {{Lattice Gauge Theory for Physics Beyond
  the Standard Model}},}\ }\href {\doibase 10.1140/epja/i2019-12901-5}
  {\bibfield  {journal} {\bibinfo  {journal} {Eur. Phys. J. A}\ }\textbf
  {\bibinfo {volume} {55}},\ \bibinfo {pages} {198} (\bibinfo {year} {2019})},\
  \Eprint {http://arxiv.org/abs/1904.09964} {arXiv:1904.09964} \BibitemShut
  {NoStop}%
\bibitem [{\citenamefont {Buarque~Franzosi}\ \emph {et~al.}(2020)\citenamefont
  {Buarque~Franzosi}, \citenamefont {Cacciapaglia},\ and\ \citenamefont
  {Deandrea}}]{BuarqueFranzosi:2018eaj}%
  \BibitemOpen
  \bibfield  {author} {\bibinfo {author} {\bibfnamefont {D.}~\bibnamefont
  {Buarque~Franzosi}}, \bibinfo {author} {\bibfnamefont {G.}~\bibnamefont
  {Cacciapaglia}}, \ and\ \bibinfo {author} {\bibfnamefont {A.}~\bibnamefont
  {Deandrea}},\ }\bibfield  {title} {\enquote {\bibinfo {title}
  {{Sigma-assisted low scale composite Goldstone--Higgs}},}\ }\href {\doibase
  10.1140/epjc/s10052-019-7572-z} {\bibfield  {journal} {\bibinfo  {journal}
  {Eur. Phys. J. C}\ }\textbf {\bibinfo {volume} {80}},\ \bibinfo {pages} {28}
  (\bibinfo {year} {2020})},\ \Eprint {http://arxiv.org/abs/1809.09146}
  {arXiv:1809.09146} \BibitemShut {NoStop}%
\bibitem [{\citenamefont {Appelquist}\ \emph
  {et~al.}(2021{\natexlab{a}})\citenamefont {Appelquist}, \citenamefont
  {Ingoldby},\ and\ \citenamefont {Piai}}]{Appelquist:2020bqj}%
  \BibitemOpen
  \bibfield  {author} {\bibinfo {author} {\bibfnamefont {T.}~\bibnamefont
  {Appelquist}}, \bibinfo {author} {\bibfnamefont {J.}~\bibnamefont
  {Ingoldby}}, \ and\ \bibinfo {author} {\bibfnamefont {M.}~\bibnamefont
  {Piai}},\ }\bibfield  {title} {\enquote {\bibinfo {title} {{Nearly Conformal
  Composite Higgs Model}},}\ }\href {\doibase 10.1103/PhysRevLett.126.191804}
  {\bibfield  {journal} {\bibinfo  {journal} {Phys. Rev. Lett.}\ }\textbf
  {\bibinfo {volume} {126}},\ \bibinfo {pages} {191804} (\bibinfo {year}
  {2021}{\natexlab{a}})},\ \Eprint {http://arxiv.org/abs/2012.09698}
  {arXiv:2012.09698} \BibitemShut {NoStop}%
\bibitem [{\citenamefont {Witzel}(2019)}]{Witzel:2019jbe}%
  \BibitemOpen
  \bibfield  {author} {\bibinfo {author} {\bibfnamefont {O.}~\bibnamefont
  {Witzel}},\ }\bibfield  {title} {\enquote {\bibinfo {title} {{Review on
  Composite Higgs Models}},}\ }\href {\doibase 10.22323/1.334.0006} {\bibfield
  {journal} {\bibinfo  {journal} {Proc. Sci.}\ }\textbf {\bibinfo {volume}
  {LATTICE2018}},\ \bibinfo {pages} {006} (\bibinfo {year} {2019})},\ \Eprint
  {http://arxiv.org/abs/1901.08216} {arXiv:1901.08216} \BibitemShut {NoStop}%
\bibitem [{\citenamefont {Appelquist}\ \emph
  {et~al.}(2021{\natexlab{b}})\citenamefont {Appelquist}, \citenamefont
  {Brower}, \citenamefont {Cushman}, \citenamefont {Fleming}, \citenamefont
  {Gasbarro}, \citenamefont {Hasenfratz}, \citenamefont {Jin}, \citenamefont
  {Neil}, \citenamefont {Osborn}, \citenamefont {Rebbi}, \citenamefont
  {Rinaldi}, \citenamefont {Schaich}, \citenamefont {Vranas},\ and\
  \citenamefont {Witzel}}]{Appelquist:2020xua}%
  \BibitemOpen
  \bibfield  {author} {\bibinfo {author} {\bibfnamefont {T.}~\bibnamefont
  {Appelquist}}, \bibinfo {author} {\bibfnamefont {R.~C.}\ \bibnamefont
  {Brower}}, \bibinfo {author} {\bibfnamefont {K.~K.}\ \bibnamefont {Cushman}},
  \bibinfo {author} {\bibfnamefont {G.~T.}\ \bibnamefont {Fleming}}, \bibinfo
  {author} {\bibfnamefont {A.~D.}\ \bibnamefont {Gasbarro}}, \bibinfo {author}
  {\bibfnamefont {A.}~\bibnamefont {Hasenfratz}}, \bibinfo {author}
  {\bibfnamefont {X.-Y.}\ \bibnamefont {Jin}}, \bibinfo {author} {\bibfnamefont
  {E.~T.}\ \bibnamefont {Neil}}, \bibinfo {author} {\bibfnamefont {J.~C.}\
  \bibnamefont {Osborn}}, \bibinfo {author} {\bibfnamefont {C.}~\bibnamefont
  {Rebbi}}, \bibinfo {author} {\bibfnamefont {E.}~\bibnamefont {Rinaldi}},
  \bibinfo {author} {\bibfnamefont {D.}~\bibnamefont {Schaich}}, \bibinfo
  {author} {\bibfnamefont {P.}~\bibnamefont {Vranas}}, \ and\ \bibinfo {author}
  {\bibfnamefont {O.}~\bibnamefont {Witzel}} (\bibinfo {collaboration} {LSD
  Collaboration}),\ }\bibfield  {title} {\enquote {\bibinfo {title}
  {{Near-conformal dynamics in a chirally broken system}},}\ }\href {\doibase
  10.1103/PhysRevD.103.014504} {\bibfield  {journal} {\bibinfo  {journal}
  {Phys. Rev. D}\ }\textbf {\bibinfo {volume} {103}},\ \bibinfo {pages}
  {014504} (\bibinfo {year} {2021}{\natexlab{b}})},\ \Eprint
  {http://arxiv.org/abs/2007.01810} {arXiv:2007.01810} \BibitemShut {NoStop}%
\bibitem [{\citenamefont {Witzel}\ \emph {et~al.}(2021)\citenamefont {Witzel},
  \citenamefont {Hasenfratz},\ and\ \citenamefont {Peterson}}]{Witzel:2020hyr}%
  \BibitemOpen
  \bibfield  {author} {\bibinfo {author} {\bibfnamefont {O.}~\bibnamefont
  {Witzel}}, \bibinfo {author} {\bibfnamefont {A.}~\bibnamefont {Hasenfratz}},
  \ and\ \bibinfo {author} {\bibfnamefont {C.~T.}\ \bibnamefont {Peterson}}
  (\bibinfo {collaboration} {LSD Collaboration}),\ }\bibfield  {title}
  {\enquote {\bibinfo {title} {{Composite Higgs scenario in mass-split
  models}},}\ }\href {\doibase 10.22323/1.390.0675} {\bibfield  {journal}
  {\bibinfo  {journal} {Proc. Sci.}\ }\textbf {\bibinfo {volume} {ICHEP2020}},\
  \bibinfo {pages} {675} (\bibinfo {year} {2021})},\ \Eprint
  {http://arxiv.org/abs/2011.05175} {arXiv:2011.05175} \BibitemShut {NoStop}%
\bibitem [{\citenamefont {Fodor}\ \emph {et~al.}(2020)\citenamefont {Fodor},
  \citenamefont {Holland}, \citenamefont {Kuti},\ and\ \citenamefont
  {Wong}}]{Fodor:2020niv}%
  \BibitemOpen
  \bibfield  {author} {\bibinfo {author} {\bibfnamefont {Z.}~\bibnamefont
  {Fodor}}, \bibinfo {author} {\bibfnamefont {K.}~\bibnamefont {Holland}},
  \bibinfo {author} {\bibfnamefont {J.}~\bibnamefont {Kuti}}, \ and\ \bibinfo
  {author} {\bibfnamefont {C.~H.}\ \bibnamefont {Wong}},\ }\bibfield  {title}
  {\enquote {\bibinfo {title} {{Dilaton EFT from p-regime to RMT in the
  $\epsilon$-regime}},}\ }\href {\doibase 10.22323/1.363.0246} {\bibfield
  {journal} {\bibinfo  {journal} {Proc. Sci.}\ }\textbf {\bibinfo {volume}
  {LATTICE2019}},\ \bibinfo {pages} {246} (\bibinfo {year} {2020})},\ \Eprint
  {http://arxiv.org/abs/2002.05163} {arXiv:2002.05163} \BibitemShut {NoStop}%
\bibitem [{\citenamefont {Weinberg}(1966)}]{Weinberg:1966kf}%
  \BibitemOpen
  \bibfield  {author} {\bibinfo {author} {\bibfnamefont {S.}~\bibnamefont
  {Weinberg}},\ }\bibfield  {title} {\enquote {\bibinfo {title} {{Pion
  scattering lengths}},}\ }\href {\doibase 10.1103/PhysRevLett.17.616}
  {\bibfield  {journal} {\bibinfo  {journal} {Phys. Rev. Lett.}\ }\textbf
  {\bibinfo {volume} {17}},\ \bibinfo {pages} {616--621} (\bibinfo {year}
  {1966})}\BibitemShut {NoStop}%
\bibitem [{\citenamefont {Lai}(1969)}]{Lai:1969xr}%
  \BibitemOpen
  \bibfield  {author} {\bibinfo {author} {\bibfnamefont {C.~S.}\ \bibnamefont
  {Lai}},\ }\bibfield  {title} {\enquote {\bibinfo {title} {{On low-energy
  pion-pion scattering lengths}},}\ }\href {\doibase 10.1007/BF02731801}
  {\bibfield  {journal} {\bibinfo  {journal} {Nuovo Cim. A}\ }\textbf {\bibinfo
  {volume} {62}},\ \bibinfo {pages} {192--198} (\bibinfo {year}
  {1969})}\BibitemShut {NoStop}%
\bibitem [{\citenamefont {L{\"u}scher}(1991)}]{Luscher:1990ux}%
  \BibitemOpen
  \bibfield  {author} {\bibinfo {author} {\bibfnamefont {M.}~\bibnamefont
  {L{\"u}scher}},\ }\bibfield  {title} {\enquote {\bibinfo {title}
  {{Two-particle states on a torus and their relation to the scattering
  matrix}},}\ }\href {\doibase 10.1016/0550-3213(91)90366-6} {\bibfield
  {journal} {\bibinfo  {journal} {Nucl. Phys. B}\ }\textbf {\bibinfo {volume}
  {354}},\ \bibinfo {pages} {531--578} (\bibinfo {year} {1991})}\BibitemShut
  {NoStop}%
\bibitem [{\citenamefont {Golterman}\ and\ \citenamefont
  {Shamir}(2016)}]{Golterman:2016lsd}%
  \BibitemOpen
  \bibfield  {author} {\bibinfo {author} {\bibfnamefont {M.}~\bibnamefont
  {Golterman}}\ and\ \bibinfo {author} {\bibfnamefont {Y.}~\bibnamefont
  {Shamir}},\ }\bibfield  {title} {\enquote {\bibinfo {title} {{Low-energy
  effective action for pions and a dilatonic meson}},}\ }\href {\doibase
  10.1103/PhysRevD.94.054502} {\bibfield  {journal} {\bibinfo  {journal} {Phys.
  Rev. D}\ }\textbf {\bibinfo {volume} {94}},\ \bibinfo {pages} {054502}
  (\bibinfo {year} {2016})},\ \Eprint {http://arxiv.org/abs/1603.04575}
  {arXiv:1603.04575} \BibitemShut {NoStop}%
\bibitem [{\citenamefont {Appelquist}\ \emph {et~al.}(2017)\citenamefont
  {Appelquist}, \citenamefont {Ingoldby},\ and\ \citenamefont
  {Piai}}]{Appelquist:2017wcg}%
  \BibitemOpen
  \bibfield  {author} {\bibinfo {author} {\bibfnamefont {Thomas}\ \bibnamefont
  {Appelquist}}, \bibinfo {author} {\bibfnamefont {James}\ \bibnamefont
  {Ingoldby}}, \ and\ \bibinfo {author} {\bibfnamefont {Maurizio}\ \bibnamefont
  {Piai}},\ }\bibfield  {title} {\enquote {\bibinfo {title} {{Dilaton EFT
  Framework For Lattice Data}},}\ }\href {\doibase 10.1007/JHEP07(2017)035}
  {\bibfield  {journal} {\bibinfo  {journal} {JHEP}\ }\textbf {\bibinfo
  {volume} {07}},\ \bibinfo {pages} {035} (\bibinfo {year} {2017})},\ \Eprint
  {http://arxiv.org/abs/1702.04410} {arXiv:1702.04410 [hep-ph]} \BibitemShut
  {NoStop}%
\bibitem [{\citenamefont {Appelquist}\ \emph
  {et~al.}(2018{\natexlab{a}})\citenamefont {Appelquist}, \citenamefont
  {Ingoldby},\ and\ \citenamefont {Piai}}]{Appelquist:2017vyy}%
  \BibitemOpen
  \bibfield  {author} {\bibinfo {author} {\bibfnamefont {Thomas}\ \bibnamefont
  {Appelquist}}, \bibinfo {author} {\bibfnamefont {James}\ \bibnamefont
  {Ingoldby}}, \ and\ \bibinfo {author} {\bibfnamefont {Maurizio}\ \bibnamefont
  {Piai}},\ }\bibfield  {title} {\enquote {\bibinfo {title} {{Analysis of a
  Dilaton EFT for Lattice Data}},}\ }\href {\doibase 10.1007/JHEP03(2018)039}
  {\bibfield  {journal} {\bibinfo  {journal} {JHEP}\ }\textbf {\bibinfo
  {volume} {03}},\ \bibinfo {pages} {039} (\bibinfo {year}
  {2018}{\natexlab{a}})},\ \Eprint {http://arxiv.org/abs/1711.00067}
  {arXiv:1711.00067 [hep-ph]} \BibitemShut {NoStop}%
\bibitem [{\citenamefont {Appelquist}\ \emph {et~al.}(2020)\citenamefont
  {Appelquist}, \citenamefont {Ingoldby},\ and\ \citenamefont
  {Piai}}]{Appelquist:2019lgk}%
  \BibitemOpen
  \bibfield  {author} {\bibinfo {author} {\bibfnamefont {T.}~\bibnamefont
  {Appelquist}}, \bibinfo {author} {\bibfnamefont {J.}~\bibnamefont
  {Ingoldby}}, \ and\ \bibinfo {author} {\bibfnamefont {M.}~\bibnamefont
  {Piai}},\ }\bibfield  {title} {\enquote {\bibinfo {title} {{Dilaton potential
  and lattice data}},}\ }\href {\doibase 10.1103/PhysRevD.101.075025}
  {\bibfield  {journal} {\bibinfo  {journal} {Phys. Rev. D}\ }\textbf {\bibinfo
  {volume} {101}},\ \bibinfo {pages} {075025} (\bibinfo {year} {2020})},\
  \Eprint {http://arxiv.org/abs/1908.00895} {arXiv:1908.00895} \BibitemShut
  {NoStop}%
\bibitem [{\citenamefont {Appelquist}\ \emph
  {et~al.}(2018{\natexlab{b}})\citenamefont {Appelquist} \emph
  {et~al.}}]{Appelquist:2018tyt}%
  \BibitemOpen
  \bibfield  {author} {\bibinfo {author} {\bibfnamefont {T.}~\bibnamefont
  {Appelquist}} \emph {et~al.} (\bibinfo {collaboration} {LSD}),\ }\bibfield
  {title} {\enquote {\bibinfo {title} {{Linear Sigma EFT for Nearly Conformal
  Gauge Theories}},}\ }\href {\doibase 10.1103/PhysRevD.98.114510} {\bibfield
  {journal} {\bibinfo  {journal} {Phys. Rev. D}\ }\textbf {\bibinfo {volume}
  {98}},\ \bibinfo {pages} {114510} (\bibinfo {year} {2018}{\natexlab{b}})},\
  \Eprint {http://arxiv.org/abs/1809.02624} {arXiv:1809.02624 [hep-ph]}
  \BibitemShut {NoStop}%
\bibitem [{\citenamefont {Hansen}\ \emph {et~al.}(2017)\citenamefont {Hansen},
  \citenamefont {Lang\ae{}ble},\ and\ \citenamefont
  {Sannino}}]{Hansen:2016fri}%
  \BibitemOpen
  \bibfield  {author} {\bibinfo {author} {\bibfnamefont {Martin}\ \bibnamefont
  {Hansen}}, \bibinfo {author} {\bibfnamefont {Kasper}\ \bibnamefont
  {Lang\ae{}ble}}, \ and\ \bibinfo {author} {\bibfnamefont {Francesco}\
  \bibnamefont {Sannino}},\ }\bibfield  {title} {\enquote {\bibinfo {title}
  {{Extending Chiral Perturbation Theory with an Isosinglet Scalar}},}\ }\href
  {\doibase 10.1103/PhysRevD.95.036005} {\bibfield  {journal} {\bibinfo
  {journal} {Phys. Rev. D}\ }\textbf {\bibinfo {volume} {95}},\ \bibinfo
  {pages} {036005} (\bibinfo {year} {2017})},\ \Eprint
  {http://arxiv.org/abs/1610.02904} {arXiv:1610.02904 [hep-ph]} \BibitemShut
  {NoStop}%
\bibitem [{\citenamefont {Hasenfratz}\ and\ \citenamefont
  {Knechtli}(2001)}]{Hasenfratz:2001hp}%
  \BibitemOpen
  \bibfield  {author} {\bibinfo {author} {\bibfnamefont {A.}~\bibnamefont
  {Hasenfratz}}\ and\ \bibinfo {author} {\bibfnamefont {F.}~\bibnamefont
  {Knechtli}},\ }\bibfield  {title} {\enquote {\bibinfo {title} {{Flavor
  symmetry and the static potential with hypercubic blocking}},}\ }\href
  {\doibase 10.1103/PhysRevD.64.034504} {\bibfield  {journal} {\bibinfo
  {journal} {Phys. Rev. D}\ }\textbf {\bibinfo {volume} {64}},\ \bibinfo
  {pages} {034504} (\bibinfo {year} {2001})},\ \Eprint
  {http://arxiv.org/abs/hep-lat/0103029} {hep-lat/0103029} \BibitemShut
  {NoStop}%
\bibitem [{\citenamefont {Hasenfratz}\ \emph {et~al.}(2007)\citenamefont
  {Hasenfratz}, \citenamefont {Hoffmann},\ and\ \citenamefont
  {Schaefer}}]{Hasenfratz:2007rf}%
  \BibitemOpen
  \bibfield  {author} {\bibinfo {author} {\bibfnamefont {A.}~\bibnamefont
  {Hasenfratz}}, \bibinfo {author} {\bibfnamefont {R.}~\bibnamefont
  {Hoffmann}}, \ and\ \bibinfo {author} {\bibfnamefont {S.}~\bibnamefont
  {Schaefer}},\ }\bibfield  {title} {\enquote {\bibinfo {title} {{Hypercubic
  smeared links for dynamical fermions}},}\ }\href {\doibase
  10.1088/1126-6708/2007/05/029} {\bibfield  {journal} {\bibinfo  {journal}
  {JHEP}\ }\textbf {\bibinfo {volume} {0705}},\ \bibinfo {pages} {029}
  (\bibinfo {year} {2007})},\ \Eprint {http://arxiv.org/abs/hep-lat/0702028}
  {hep-lat/0702028} \BibitemShut {NoStop}%
\bibitem [{\citenamefont {Cheng}\ \emph {et~al.}(2012)\citenamefont {Cheng},
  \citenamefont {Hasenfratz},\ and\ \citenamefont {Schaich}}]{Cheng:2011ic}%
  \BibitemOpen
  \bibfield  {author} {\bibinfo {author} {\bibfnamefont {A.}~\bibnamefont
  {Cheng}}, \bibinfo {author} {\bibfnamefont {A.}~\bibnamefont {Hasenfratz}}, \
  and\ \bibinfo {author} {\bibfnamefont {D.}~\bibnamefont {Schaich}},\
  }\bibfield  {title} {\enquote {\bibinfo {title} {{Novel phase in SU(3)
  lattice gauge theory with 12 light fermions}},}\ }\href {\doibase
  10.1103/PhysRevD.85.094509} {\bibfield  {journal} {\bibinfo  {journal} {Phys.
  Rev. D}\ }\textbf {\bibinfo {volume} {85}},\ \bibinfo {pages} {094509}
  (\bibinfo {year} {2012})},\ \Eprint {http://arxiv.org/abs/1111.2317}
  {arXiv:1111.2317} \BibitemShut {NoStop}%
\bibitem [{\citenamefont {Schaich}\ \emph {et~al.}(2015)\citenamefont
  {Schaich}, \citenamefont {Hasenfratz},\ and\ \citenamefont
  {Rinaldi}}]{Schaich:2015psa}%
  \BibitemOpen
  \bibfield  {author} {\bibinfo {author} {\bibfnamefont {David}\ \bibnamefont
  {Schaich}}, \bibinfo {author} {\bibfnamefont {Anna}\ \bibnamefont
  {Hasenfratz}}, \ and\ \bibinfo {author} {\bibfnamefont {Enrico}\ \bibnamefont
  {Rinaldi}} (\bibinfo {collaboration} {LSD}),\ }\bibfield  {title} {\enquote
  {\bibinfo {title} {{Finite-temperature study of eight-flavor SU(3) gauge
  theory}},}\ }in\ \href {\doibase 10.1142/9789813231467_0051} {\emph {\bibinfo
  {booktitle} {{Sakata Memorial KMI Workshop on Origin of Mass and Strong
  Coupling Gauge Theories}}}}\ (\bibinfo  {publisher} {WSP},\ \bibinfo {year}
  {2015})\ \Eprint {http://arxiv.org/abs/1506.08791} {arXiv:1506.08791
  [hep-lat]} \BibitemShut {NoStop}%
\bibitem [{\citenamefont {Daniel}\ and\ \citenamefont
  {Sheard}(1988)}]{Daniel:1987aa}%
  \BibitemOpen
  \bibfield  {author} {\bibinfo {author} {\bibfnamefont {D.}~\bibnamefont
  {Daniel}}\ and\ \bibinfo {author} {\bibfnamefont {S.~N.}\ \bibnamefont
  {Sheard}},\ }\bibfield  {title} {\enquote {\bibinfo {title} {{Perturbative
  corrections to staggered-fermion lattice operators}},}\ }\href {\doibase
  10.1016/0550-3213(88)90211-8} {\bibfield  {journal} {\bibinfo  {journal}
  {Nucl. Phys. B}\ }\textbf {\bibinfo {volume} {302}},\ \bibinfo {pages}
  {471--498} (\bibinfo {year} {1988})}\BibitemShut {NoStop}%
\bibitem [{\citenamefont {Kilcup}\ and\ \citenamefont
  {Sharpe}(1987)}]{Kilcup:1986dg}%
  \BibitemOpen
  \bibfield  {author} {\bibinfo {author} {\bibfnamefont {G.~W.}\ \bibnamefont
  {Kilcup}}\ and\ \bibinfo {author} {\bibfnamefont {Stephen~R.}\ \bibnamefont
  {Sharpe}},\ }\bibfield  {title} {\enquote {\bibinfo {title} {{A Tool Kit for
  Staggered Fermions}},}\ }\href {\doibase 10.1016/0550-3213(87)90285-9}
  {\bibfield  {journal} {\bibinfo  {journal} {Nucl. Phys. B}\ }\textbf
  {\bibinfo {volume} {283}},\ \bibinfo {pages} {493--550} (\bibinfo {year}
  {1987})}\BibitemShut {NoStop}%
\bibitem [{\citenamefont {Sharpe}\ and\ \citenamefont {Van~de
  Water}(2005)}]{Sharpe:2004is}%
  \BibitemOpen
  \bibfield  {author} {\bibinfo {author} {\bibfnamefont {S.~R.}\ \bibnamefont
  {Sharpe}}\ and\ \bibinfo {author} {\bibfnamefont {R.~S.}\ \bibnamefont
  {Van~de Water}},\ }\bibfield  {title} {\enquote {\bibinfo {title} {{Staggered
  chiral perturbation theory at next-to-leading order}},}\ }\href {\doibase
  10.1103/PhysRevD.71.114505} {\bibfield  {journal} {\bibinfo  {journal} {Phys.
  Rev. D}\ }\textbf {\bibinfo {volume} {71}},\ \bibinfo {pages} {114505}
  (\bibinfo {year} {2005})},\ \Eprint {http://arxiv.org/abs/hep-lat/0409018}
  {hep-lat/0409018} \BibitemShut {NoStop}%
\bibitem [{\citenamefont {Kuramashi}\ \emph {et~al.}(1993)\citenamefont
  {Kuramashi}, \citenamefont {Fukugita}, \citenamefont {Mino}, \citenamefont
  {Okawa},\ and\ \citenamefont {Ukawa}}]{Kuramashi:1993ka}%
  \BibitemOpen
  \bibfield  {author} {\bibinfo {author} {\bibfnamefont {Y.}~\bibnamefont
  {Kuramashi}}, \bibinfo {author} {\bibfnamefont {M.}~\bibnamefont {Fukugita}},
  \bibinfo {author} {\bibfnamefont {H.}~\bibnamefont {Mino}}, \bibinfo {author}
  {\bibfnamefont {M.}~\bibnamefont {Okawa}}, \ and\ \bibinfo {author}
  {\bibfnamefont {A.}~\bibnamefont {Ukawa}},\ }\bibfield  {title} {\enquote
  {\bibinfo {title} {{Lattice QCD calculation of full pion scattering
  lengths}},}\ }\href {\doibase 10.1103/PhysRevLett.71.2387} {\bibfield
  {journal} {\bibinfo  {journal} {Phys. Rev. Lett.}\ }\textbf {\bibinfo
  {volume} {71}},\ \bibinfo {pages} {2387--2390} (\bibinfo {year}
  {1993})}\BibitemShut {NoStop}%
\bibitem [{\citenamefont {Fukugita}\ \emph {et~al.}(1995)\citenamefont
  {Fukugita}, \citenamefont {Kuramashi}, \citenamefont {Okawa}, \citenamefont
  {Mino},\ and\ \citenamefont {Ukawa}}]{Fukugita:1994ve}%
  \BibitemOpen
  \bibfield  {author} {\bibinfo {author} {\bibfnamefont {M.}~\bibnamefont
  {Fukugita}}, \bibinfo {author} {\bibfnamefont {Y.}~\bibnamefont {Kuramashi}},
  \bibinfo {author} {\bibfnamefont {M.}~\bibnamefont {Okawa}}, \bibinfo
  {author} {\bibfnamefont {H.}~\bibnamefont {Mino}}, \ and\ \bibinfo {author}
  {\bibfnamefont {A.}~\bibnamefont {Ukawa}},\ }\bibfield  {title} {\enquote
  {\bibinfo {title} {{Hadron scattering lengths in lattice QCD}},}\ }\href
  {\doibase 10.1103/PhysRevD.52.3003} {\bibfield  {journal} {\bibinfo
  {journal} {Phys. Rev. D}\ }\textbf {\bibinfo {volume} {52}},\ \bibinfo
  {pages} {3003--3023} (\bibinfo {year} {1995})},\ \Eprint
  {http://arxiv.org/abs/hep-lat/9501024} {hep-lat/9501024} \BibitemShut
  {NoStop}%
\bibitem [{\citenamefont {Fu}(2013)}]{Fu:2013ffa}%
  \BibitemOpen
  \bibfield  {author} {\bibinfo {author} {\bibfnamefont {Z.}~\bibnamefont
  {Fu}},\ }\bibfield  {title} {\enquote {\bibinfo {title} {{Lattice QCD study
  of the s-wave $\pi\pi $ scattering lengths in the I=0 and 2 channels}},}\
  }\href {\doibase 10.1103/PhysRevD.87.074501} {\bibfield  {journal} {\bibinfo
  {journal} {Phys. Rev. D}\ }\textbf {\bibinfo {volume} {87}},\ \bibinfo
  {pages} {074501} (\bibinfo {year} {2013})},\ \Eprint
  {http://arxiv.org/abs/1303.0517} {arXiv:1303.0517} \BibitemShut {NoStop}%
\bibitem [{\citenamefont {Akaike}(1974)}]{Akaike:1974AIC}%
  \BibitemOpen
  \bibfield  {author} {\bibinfo {author} {\bibfnamefont {H.}~\bibnamefont
  {Akaike}},\ }\bibfield  {title} {\enquote {\bibinfo {title} {{A new look at
  the statistical model identification}},}\ }\href {\doibase
  10.1109/TAC.1974.1100705} {\bibfield  {journal} {\bibinfo  {journal} {IEEE
  Trans. Autom. Control}\ }\textbf {\bibinfo {volume} {19}},\ \bibinfo {pages}
  {716--723} (\bibinfo {year} {1974})}\BibitemShut {NoStop}%
\bibitem [{\citenamefont {Jay}\ and\ \citenamefont {Neil}(2021)}]{Jay:2020jkz}%
  \BibitemOpen
  \bibfield  {author} {\bibinfo {author} {\bibfnamefont {W.~I.}\ \bibnamefont
  {Jay}}\ and\ \bibinfo {author} {\bibfnamefont {E.~T.}\ \bibnamefont {Neil}},\
  }\bibfield  {title} {\enquote {\bibinfo {title} {{Bayesian model averaging
  for analysis of lattice field theory results}},}\ }\href {\doibase
  10.1103/PhysRevD.103.114502} {\bibfield  {journal} {\bibinfo  {journal}
  {Phys. Rev. D}\ }\textbf {\bibinfo {volume} {103}},\ \bibinfo {pages}
  {114502} (\bibinfo {year} {2021})},\ \Eprint
  {http://arxiv.org/abs/2008.01069} {arXiv:2008.01069} \BibitemShut {NoStop}%
\bibitem [{\citenamefont {Gasser}\ and\ \citenamefont
  {Leutwyler}(1987)}]{Gasser:1986vb}%
  \BibitemOpen
  \bibfield  {author} {\bibinfo {author} {\bibfnamefont {J.}~\bibnamefont
  {Gasser}}\ and\ \bibinfo {author} {\bibfnamefont {H.}~\bibnamefont
  {Leutwyler}},\ }\bibfield  {title} {\enquote {\bibinfo {title} {{Light Quarks
  at Low Temperatures}},}\ }\href {\doibase 10.1016/0370-2693(87)90492-8}
  {\bibfield  {journal} {\bibinfo  {journal} {Phys. Lett. B}\ }\textbf
  {\bibinfo {volume} {184}},\ \bibinfo {pages} {83--88} (\bibinfo {year}
  {1987})}\BibitemShut {NoStop}%
\bibitem [{\citenamefont {Leutwyler}(1987)}]{Leutwyler:1987ak}%
  \BibitemOpen
  \bibfield  {author} {\bibinfo {author} {\bibfnamefont {H.}~\bibnamefont
  {Leutwyler}},\ }\bibfield  {title} {\enquote {\bibinfo {title} {{Energy
  Levels of Light Quarks Confined to a Box}},}\ }\href {\doibase
  10.1016/0370-2693(87)91296-2} {\bibfield  {journal} {\bibinfo  {journal}
  {Phys. Lett. B}\ }\textbf {\bibinfo {volume} {189}},\ \bibinfo {pages}
  {197--202} (\bibinfo {year} {1987})}\BibitemShut {NoStop}%
\bibitem [{\citenamefont {Gasser}\ and\ \citenamefont
  {Leutwyler}(1988)}]{Gasser:1987zq}%
  \BibitemOpen
  \bibfield  {author} {\bibinfo {author} {\bibfnamefont {J.}~\bibnamefont
  {Gasser}}\ and\ \bibinfo {author} {\bibfnamefont {H.}~\bibnamefont
  {Leutwyler}},\ }\bibfield  {title} {\enquote {\bibinfo {title}
  {{Spontaneously Broken Symmetries: Effective Lagrangians at Finite
  Volume}},}\ }\href {\doibase 10.1016/0550-3213(88)90107-1} {\bibfield
  {journal} {\bibinfo  {journal} {Nucl. Phys. B}\ }\textbf {\bibinfo {volume}
  {307}},\ \bibinfo {pages} {763--778} (\bibinfo {year} {1988})}\BibitemShut
  {NoStop}%
\bibitem [{\citenamefont {L{\"u}scher}(1986)}]{Luscher:1985dn}%
  \BibitemOpen
  \bibfield  {author} {\bibinfo {author} {\bibfnamefont {M.}~\bibnamefont
  {L{\"u}scher}},\ }\bibfield  {title} {\enquote {\bibinfo {title} {{Volume
  Dependence of the Energy Spectrum in Massive Quantum Field Theories. 1.
  Stable Particle States}},}\ }\href {\doibase 10.1007/BF01211589} {\bibfield
  {journal} {\bibinfo  {journal} {Commun. Math. Phys.}\ }\textbf {\bibinfo
  {volume} {104}},\ \bibinfo {pages} {177} (\bibinfo {year}
  {1986})}\BibitemShut {NoStop}%
\bibitem [{\citenamefont {Colangelo}\ and\ \citenamefont
  {Durr}(2004)}]{Colangelo:2003hf}%
  \BibitemOpen
  \bibfield  {author} {\bibinfo {author} {\bibfnamefont {G.}~\bibnamefont
  {Colangelo}}\ and\ \bibinfo {author} {\bibfnamefont {S.}~\bibnamefont
  {Durr}},\ }\bibfield  {title} {\enquote {\bibinfo {title} {{The Pion mass in
  finite volume}},}\ }\href {\doibase 10.1140/epjc/s2004-01593-y} {\bibfield
  {journal} {\bibinfo  {journal} {Eur. Phys. J. C}\ }\textbf {\bibinfo {volume}
  {33}},\ \bibinfo {pages} {543--553} (\bibinfo {year} {2004})},\ \Eprint
  {http://arxiv.org/abs/hep-lat/0311023} {hep-lat/0311023} \BibitemShut
  {NoStop}%
\bibitem [{\citenamefont {Colangelo}\ and\ \citenamefont
  {Haefeli}(2004)}]{Colangelo:2004xr}%
  \BibitemOpen
  \bibfield  {author} {\bibinfo {author} {\bibfnamefont {G.}~\bibnamefont
  {Colangelo}}\ and\ \bibinfo {author} {\bibfnamefont {C.}~\bibnamefont
  {Haefeli}},\ }\bibfield  {title} {\enquote {\bibinfo {title} {{An Asymptotic
  formula for the pion decay constant in a large volume}},}\ }\href {\doibase
  10.1016/j.physletb.2004.03.080} {\bibfield  {journal} {\bibinfo  {journal}
  {Phys. Lett. B}\ }\textbf {\bibinfo {volume} {590}},\ \bibinfo {pages}
  {258--264} (\bibinfo {year} {2004})},\ \Eprint
  {http://arxiv.org/abs/hep-lat/0403025} {hep-lat/0403025} \BibitemShut
  {NoStop}%
\bibitem [{\citenamefont {Leung}\ \emph {et~al.}(1989)\citenamefont {Leung},
  \citenamefont {Love},\ and\ \citenamefont {Bardeen}}]{Leung:1989hw}%
  \BibitemOpen
  \bibfield  {author} {\bibinfo {author} {\bibfnamefont {C.~N.}\ \bibnamefont
  {Leung}}, \bibinfo {author} {\bibfnamefont {S.~T.}\ \bibnamefont {Love}}, \
  and\ \bibinfo {author} {\bibfnamefont {W.~A.}\ \bibnamefont {Bardeen}},\
  }\bibfield  {title} {\enquote {\bibinfo {title} {{Aspects of Dynamical
  Symmetry Breaking in Gauge Field Theories}},}\ }\href {\doibase
  10.1016/0550-3213(89)90121-1} {\bibfield  {journal} {\bibinfo  {journal}
  {Nucl. Phys. B}\ }\textbf {\bibinfo {volume} {323}},\ \bibinfo {pages}
  {493--512} (\bibinfo {year} {1989})}\BibitemShut {NoStop}%
\bibitem [{\citenamefont {Chacko}\ and\ \citenamefont
  {Mishra}(2013)}]{Chacko:2012sy}%
  \BibitemOpen
  \bibfield  {author} {\bibinfo {author} {\bibfnamefont {Z.}~\bibnamefont
  {Chacko}}\ and\ \bibinfo {author} {\bibfnamefont {R.~K.}\ \bibnamefont
  {Mishra}},\ }\bibfield  {title} {\enquote {\bibinfo {title} {{Effective
  Theory of a Light Dilaton}},}\ }\href {\doibase 10.1103/PhysRevD.87.115006}
  {\bibfield  {journal} {\bibinfo  {journal} {Phys. Rev. D}\ }\textbf {\bibinfo
  {volume} {87}},\ \bibinfo {pages} {115006} (\bibinfo {year} {2013})},\
  \Eprint {http://arxiv.org/abs/1209.3022} {arXiv:1209.3022} \BibitemShut
  {NoStop}%
\bibitem [{\citenamefont {Cat\`{a}}\ \emph {et~al.}(2019)\citenamefont
  {Cat\`{a}}, \citenamefont {Crewther},\ and\ \citenamefont
  {Tunstall}}]{Cata:2018wzl}%
  \BibitemOpen
  \bibfield  {author} {\bibinfo {author} {\bibfnamefont {O.}~\bibnamefont
  {Cat\`{a}}}, \bibinfo {author} {\bibfnamefont {R.~J.}\ \bibnamefont
  {Crewther}}, \ and\ \bibinfo {author} {\bibfnamefont {L.~C.}\ \bibnamefont
  {Tunstall}},\ }\bibfield  {title} {\enquote {\bibinfo {title} {{Crawling
  technicolor}},}\ }\href {\doibase 10.1103/PhysRevD.100.095007} {\bibfield
  {journal} {\bibinfo  {journal} {Phys. Rev. D}\ }\textbf {\bibinfo {volume}
  {100}},\ \bibinfo {pages} {095007} (\bibinfo {year} {2019})},\ \Eprint
  {http://arxiv.org/abs/1803.08513} {arXiv:1803.08513} \BibitemShut {NoStop}%
\bibitem [{\citenamefont {Appelquist}\ and\ \citenamefont
  {Bai}(2010)}]{Appelquist:2010gy}%
  \BibitemOpen
  \bibfield  {author} {\bibinfo {author} {\bibfnamefont {T.}~\bibnamefont
  {Appelquist}}\ and\ \bibinfo {author} {\bibfnamefont {Y.}~\bibnamefont
  {Bai}},\ }\bibfield  {title} {\enquote {\bibinfo {title} {{A Light Dilaton in
  Walking Gauge Theories}},}\ }\href {\doibase 10.1103/PhysRevD.82.071701}
  {\bibfield  {journal} {\bibinfo  {journal} {Phys. Rev. D}\ }\textbf {\bibinfo
  {volume} {82}},\ \bibinfo {pages} {071701} (\bibinfo {year} {2010})},\
  \Eprint {http://arxiv.org/abs/1006.4375} {arXiv:1006.4375} \BibitemShut
  {NoStop}%
\bibitem [{\citenamefont {Golterman}\ \emph {et~al.}(2020)\citenamefont
  {Golterman}, \citenamefont {Neil},\ and\ \citenamefont
  {Shamir}}]{Golterman:2020tdq}%
  \BibitemOpen
  \bibfield  {author} {\bibinfo {author} {\bibfnamefont {M.}~\bibnamefont
  {Golterman}}, \bibinfo {author} {\bibfnamefont {E.~T.}\ \bibnamefont {Neil}},
  \ and\ \bibinfo {author} {\bibfnamefont {Y.}~\bibnamefont {Shamir}},\
  }\bibfield  {title} {\enquote {\bibinfo {title} {{Application of dilaton
  chiral perturbation theory to $N_f=8$, SU(3) spectral data}},}\ }\href
  {\doibase 10.1103/PhysRevD.102.034515} {\bibfield  {journal} {\bibinfo
  {journal} {Phys. Rev. D}\ }\textbf {\bibinfo {volume} {102}},\ \bibinfo
  {pages} {034515} (\bibinfo {year} {2020})},\ \Eprint
  {http://arxiv.org/abs/2003.00114} {arXiv:2003.00114} \BibitemShut {NoStop}%
\bibitem [{\citenamefont {Golterman}\ and\ \citenamefont
  {Shamir}(2020)}]{Golterman:2020utm}%
  \BibitemOpen
  \bibfield  {author} {\bibinfo {author} {\bibfnamefont {Maarten}\ \bibnamefont
  {Golterman}}\ and\ \bibinfo {author} {\bibfnamefont {Yigal}\ \bibnamefont
  {Shamir}},\ }\bibfield  {title} {\enquote {\bibinfo {title} {{Explorations
  beyond dilaton chiral perturbation theory in the eight-flavor SU(3) gauge
  theory}},}\ }\href {\doibase 10.1103/PhysRevD.102.114507} {\bibfield
  {journal} {\bibinfo  {journal} {Phys. Rev. D}\ }\textbf {\bibinfo {volume}
  {102}},\ \bibinfo {pages} {114507} (\bibinfo {year} {2020})},\ \Eprint
  {http://arxiv.org/abs/2009.13846} {arXiv:2009.13846 [hep-lat]} \BibitemShut
  {NoStop}%
\bibitem [{\citenamefont {Goldberger}\ \emph {et~al.}(2008)\citenamefont
  {Goldberger}, \citenamefont {Grinstein},\ and\ \citenamefont
  {Skiba}}]{Goldberger:2007zk}%
  \BibitemOpen
  \bibfield  {author} {\bibinfo {author} {\bibfnamefont {Walter~D.}\
  \bibnamefont {Goldberger}}, \bibinfo {author} {\bibfnamefont {Benjamin}\
  \bibnamefont {Grinstein}}, \ and\ \bibinfo {author} {\bibfnamefont {Witold}\
  \bibnamefont {Skiba}},\ }\bibfield  {title} {\enquote {\bibinfo {title}
  {{Distinguishing the Higgs boson from the dilaton at the Large Hadron
  Collider}},}\ }\href {\doibase 10.1103/PhysRevLett.100.111802} {\bibfield
  {journal} {\bibinfo  {journal} {Phys. Rev. Lett.}\ }\textbf {\bibinfo
  {volume} {100}},\ \bibinfo {pages} {111802} (\bibinfo {year} {2008})},\
  \Eprint {http://arxiv.org/abs/0708.1463} {arXiv:0708.1463 [hep-ph]}
  \BibitemShut {NoStop}%
\bibitem [{\citenamefont {Appelquist}\ \emph {et~al.}(2012)\citenamefont
  {Appelquist}, \citenamefont {Babich}, \citenamefont {Brower}, \citenamefont
  {Buchoff}, \citenamefont {Cheng}, \citenamefont {Clark}, \citenamefont
  {Cohen}, \citenamefont {Fleming}, \citenamefont {Kiskis}, \citenamefont
  {Lin}, \citenamefont {Neil}, \citenamefont {Osborn}, \citenamefont {Rebbi},
  \citenamefont {Schaich}, \citenamefont {Syritsyn}, \citenamefont {Voronov},
  \citenamefont {Vranas},\ and\ \citenamefont {Wasem}}]{Appelquist:2012sm}%
  \BibitemOpen
  \bibfield  {author} {\bibinfo {author} {\bibfnamefont {T.}~\bibnamefont
  {Appelquist}}, \bibinfo {author} {\bibfnamefont {R.}~\bibnamefont {Babich}},
  \bibinfo {author} {\bibfnamefont {R.~C.}\ \bibnamefont {Brower}}, \bibinfo
  {author} {\bibfnamefont {M.~I.}\ \bibnamefont {Buchoff}}, \bibinfo {author}
  {\bibfnamefont {M.}~\bibnamefont {Cheng}}, \bibinfo {author} {\bibfnamefont
  {M.~A.}\ \bibnamefont {Clark}}, \bibinfo {author} {\bibfnamefont {S.~D.}\
  \bibnamefont {Cohen}}, \bibinfo {author} {\bibfnamefont {G.~T.}\ \bibnamefont
  {Fleming}}, \bibinfo {author} {\bibfnamefont {J.}~\bibnamefont {Kiskis}},
  \bibinfo {author} {\bibfnamefont {M.}~\bibnamefont {Lin}}, \bibinfo {author}
  {\bibfnamefont {E.~T.}\ \bibnamefont {Neil}}, \bibinfo {author}
  {\bibfnamefont {J.~C.}\ \bibnamefont {Osborn}}, \bibinfo {author}
  {\bibfnamefont {C.}~\bibnamefont {Rebbi}}, \bibinfo {author} {\bibfnamefont
  {D.}~\bibnamefont {Schaich}}, \bibinfo {author} {\bibfnamefont
  {S.}~\bibnamefont {Syritsyn}}, \bibinfo {author} {\bibfnamefont
  {G.}~\bibnamefont {Voronov}}, \bibinfo {author} {\bibfnamefont
  {P.}~\bibnamefont {Vranas}}, \ and\ \bibinfo {author} {\bibfnamefont
  {J.}~\bibnamefont {Wasem}} (\bibinfo {collaboration} {LSD Collaboration}),\
  }\bibfield  {title} {\enquote {\bibinfo {title} {{WW Scattering Parameters
  via Pseudoscalar Phase Shifts}},}\ }\href {\doibase
  10.1103/PhysRevD.85.074505} {\bibfield  {journal} {\bibinfo  {journal} {Phys.
  Rev. D}\ }\textbf {\bibinfo {volume} {85}},\ \bibinfo {pages} {074505}
  (\bibinfo {year} {2012})},\ \Eprint {http://arxiv.org/abs/1201.3977}
  {arXiv:1201.3977} \BibitemShut {NoStop}%
\end{thebibliography}%
\end{document}